%% file: DOS.tex
\documentstyle[12pt,epsfig]{ioplppt}
\input{macro}

\begin{document}

\input{titlepage}

\input{intro}

\input{solution}

\input{moments}
\input{brezin}
\input{rho}
\input{higher}
\input{discussion}
\input{acknowl}

\appendix
\input{appA}
\input{appC}

\pagebreak[4]

\input{bibliography}
\end{document}

%% file: macro.tex
\newcommand{\be}{\begin{equation}}
\newcommand{\ee}{\end{equation}}
\newcommand{\bea}{\begin{eqnarray}}
\newcommand{\eea}{\end{eqnarray}}

\newcommand{\cP}{{\cal P}}

\newcommand{\vr}{{\vec r}}
\newcommand{\vrho}{{\vec \rho}}
\newcommand{\itot}{\int_{-\infty}^{\infty}}
\newcommand{\B}{{\mathrm B}}
\newcommand{\W}{{\mathrm W}}
\renewcommand{\Im}{{\mathrm Im}}

%% file: titlepage.tex
\mbox{ } \hfill cond-mat/9707153 \\
\mbox{ } \hfill Theoretical Physics Seminar in Trondheim No. 9 1997
\jl{3}

\title{Averaged Green function and density of states for
electrons in a high magnetic field and random potential}[Averaged
Green function and density of states]

\author{Anders Kristoffersen and K{\aa}re Olaussen}

\address{Institutt for fysikk, NTNU, N--7034 Trondheim, Norway}

\input{abstract}



%% file: abstract.tex
\begin{abstract}

We consider a model for $2D$ electrons
in  a very strong magnetic field
(i.e.\ projected onto a single
Landau level) and a random potential $V$.
The computation of the averaged Green
function for this system reduces to
calculating the averaged density
of states. We have constructed a
computer algebra program which
automatically generates a perturbation
expansion in $V$ for these quantities.
This is equivalent to computing moments
of the density of states.
When $V$ is a sum of Gaussians from
Poisson distributed impurities, each term in
the perturbation expansion can be evaluated
automatically. We have done so up to $12th$
order. The resulting information can be
used to reconstruct the density of states
to good precision.

\end{abstract}

%% file: intro.tex
\section{Introduction  \label{sec:intro}}

The Quantum Hall effects has continued to
furnish theoretical physics with challenges
and interesting problems for almost twenty
years. Is is remarkable that the
combination of basic physics 
(classical electrostatics and magnetism,
and non-relativistic quantum mechanics)
which has been known and researched upon
for so long can lead to so many unexpected
and exotic phenomena.

In the work reported on in this paper
we have been inspired by the integer
Quantum Hall effect\cite{Klitzing} to initiate a somewhat
prosaic (but we think nevertheless useful) project
--- the brute force high order perturbation expansion
of impurity averaged quantities in the Quantum Hall
system. The problem we consider is a model of
non-interacting electrons confined to a two-dimensional
layer in a very strong perpendicular magnetic
field and a random impurity potential $V$. All information
about the physical properties of this model
is encoded in its various Green functions.
In the absence of electron--electron interactions it
suffices to consider the one-particle Green function,
e.g.\ the solution to an equation like
\[
  -\case{\partial}{\partial t} G({\vec r},{\vec \rho};t) =
  H G({\vec r},{\vec \rho};t),
\]
where $H$ is the one-particle Hamiltonian. Thus, $G$ is a
random quantity (being a functional of $V$) for which we can
only expect to be able to compute various impurity averages like
\(
  \overline{\prod_{k=1}^K G({\vec r}_k,{\vec \rho}_k;t_k)}
\).
In this paper we shall only consider the case of $K=1$.
To investigate transport properties one must also consider the
case of $K=2$.

Our perturbation expansion consists of expanding $G$
into a series in powers of $V$ (which for our model is
equivalent to an expansion in powers of time),
and performing the impurity average
of each term in this series. The result of the averaging
process can be represented by a sum of (Feynman) diagrams.
The number of diagrams grows quite rapidly with the perturbative
order, to $12$'th order it reaches a few hundred thousand
(the exact numbers are listed in Table 1). We have written a
computer algebra program which automatically generates all
diagrams to a given order.

The use of the generated diagrams is not restricted to the
Quantum Hall system. However, the great simplification which
occurs for this system (under conditions explained below)
is that each diagram reduces to a low-dimensional Gaussian
integral (or the integral of a Gaussian multiplied by a
polynomial), and thus can be evaluated analytically by computer
algebra. The conditions are i) that the system is projected
onto a single Landau level (most simply the lowest one), and
ii) that all correlators
\(
  \overline{V({\vec r}_1) \cdots V({\vec r}_k)}
\),
can be written as a sum of Gaussians
(or Gaussians multiplied by polynomials). The latter
restriction is fulfilled if we assume $V$ to arise from
a density of Poisson distributed impurities with a Gaussian
impurity potential.
In the $\delta$-function limit, when the range of the Gaussian goes to zero,
this model belongs to a class of systems which was solved exactly for
the density of states by
Br\'{e}zin {\em et.\ al.}\cite{Brezin} (a simpler system, which
corresponds to the additional limit of taking the impurity density
to infinity, was solved earlier by Wegner\cite{Wegner}.
These solutions provide useful
checks of our results and methodology.

Given the beginning of a perturbation series, even to high orders,
it is not obvious how one should extract the correct physical
information from it, if its expansion parameter fail to be small.
It is not even obvious that computing a few more terms in the
perturbation expansion will be helpful. This is a problem
which arises quite often, and for which many methods of massaging the
perturbation expansions have been devised. Our model provide a
non-trivial example of this problem, which can be investigated to
rather high order of perturbation theory. We have considered the most
common methods of summing infinite subsets of the perturbation series.
The results were not encouraging. This is not surprising, in
view of the fact that such subsets are usually selected more by their
property of being (easily) summable than by their physical importance.

A more fruitful approach to our problem is to utilize
the fact that the information contained in the $n$'th perturbative
order is equivalent to the $n$'th moment(s) of a spectral function. For
the averaged Green function this spectral function is actually the
density of states (as function of energy).
This is due to the fact that translation invariance forces the relation
\[
   \overline{G({\vec x},{\vec y};t)} = {\cal P}({\vec x},{\vec y})\,G(t),
\]
where ${\cal P}$ is the projection operator onto the given Landau level.

There are, of course, infinitely many spectral functions which reproduce
a given finite set of moments. However, for physical systems the assumptions
of a certain degree of smoothness and simple asymptotic behaviour are usually
reasonable. Then, the high order behaviour of the moments provide information
about the tail of the spectral function, while their totality pins down its
overall behaviour. We have applied this procedure with very
satisfactory results.

The rest of this paper is organized as follows: In section 2 we give a more
detailed description of our model and the method of solution. In section 3 we
describe some of the functions which naturally occur in the graphical expansion,
the relations between them, and the number of graphs contributing to each
perturbative order. This information has been used to check the correctness
of our computer algebra programs. In section 4 we present our calculated
moments, for the lowest Landau level, and give some physical interpretation
of the results. In section 5 we discuss the exact results of Br\'{e}zin
{\em et.\ al.} and Wegner, and compare their density of states with
various ways to reconstruct it from our computed moments. In section 6
we discuss the more general case of finite range impurity potential,
still in the lowest Landau level. In section 7 we make a more restricted
analysis for some higher Landau levels. 


%% file: solution.tex
\section{Model and method of solution \label{sec:solution}}

We choose units such that the magnetic length, $\ell_B=\sqrt{\hbar/eB}$, becomes
unity. I.e., we have $(2\pi)^{-1}$ states per area in each Landau level.
We measure energies relative to the (unperturbed) energy of the Landau
level we project onto, and choose the energy scale such that the impurity potential
becomes
\begin{equation}
   \label{OneImpurityPotential}
   W({\vec r}) = \frac{2}{\varrho-1}\,
   \exp\left(\frac{-{\vec r}^{\,2}}{\varrho-1}\right),
\end{equation}
with $\varrho>1$. This is normalized to $\int \d{\vec r}\,W({\vec r})=2\pi$.
The limit $\varrho\to1^+$  corresponds to a delta function potential. 
We assume a density $f/2\pi$ of Poisson distributed impurities,
such that the random potential becomes
\begin{equation}
  V({\vec r}) = \sum_i W({\vec r}-{\vec s}_i),
\end{equation}
where the impurity positions $\{ {\vec s}_i \}$ are independently
and homogeneously distributed. I.e., there is on the average $f$
impurities per state in a single Landau level.
Since we neglect couplings to other Landau levels the
Hamiltonian is simply equal to $V$, which we shall formally view
as a perturbation. Thus, the zero'th order Green function
equals the (kernel of the) projection operator
onto the $\nu$'th Landau level,
\[
   G^{(0)}({\vec r},{\vec \rho};t) = {\cal P}_{\nu}({\vec r},{\vec \rho}).
\]
An explicit expression for this projection is
\be
  \cP_{\nu}({\vec r},{\vec \rho})=
  \frac{1}{2 \pi} \exp\left[-\case{1}{4} ({\vec r}-{\vec \rho} )^2 
  + \case{\i}{2} (x \eta - \xi y)\right]\,\times\,
   L_{\nu}\left(\case{1}{2}({\vec r}-{\vec \rho})^2\right),
  \label{eq:P}
\ee
where $(x,y)$ and $(\xi,\eta)$ are the components of
${\vec r}$ and ${\vec \rho}$ respectively, and $L_{\nu}$ is the $\nu$'th Laguerre
polynomial. For completeness we present a derivation of this well known 
result in \ref{app:P}.
Depending on the choice of gauge, the expression (\ref{eq:P}) may be
multiplied by a gauge factor
$\exp\left[\i\Gamma({\vec r})-\i\Gamma({\vec\rho})\right]$. Since $\cP_{\nu}$ is
a projection it has the reproducing property,
\be
   \int \d\vrho \; \cP(\vr_1,\vrho)\, \cP(\vrho,\vr_2) = \cP(\vr_1,\vr_2).
  \label{eq:P1}
\ee
Since $L_{\nu}(0)=1$, it follows from (\ref{eq:P}) that
$\cP_{\nu}({\vec r},{\vec r})=(2\pi)^{-1}$, in agreement with the fact that it
must equal the number of states per area.

A series expansion for the full $G$ can be generated by repeated time integrations, 
\begin{equation}
  \label{GreenFunction}
  G({\vec r},{\vec \rho};t) = \sum_{k=0}^{\infty} \frac{(-t)^k}{k!}
  \left[({\cal P}_{\nu} V)^k
  {\cal P}_{\nu} \right]({\vec r},{\vec \rho}). 
\end{equation}
We now average over the impurity positions $\{ {\vec s}_i \}$.
To this end we temporarily assume the system to be confined to
finite area $2\pi N$ with $S=f N$ impurities, and let $N\to\infty$
afterwards. We find
\begin{eqnarray}
  \label{ImpurityPotential}
   &&U({\vec r}_1,\ldots,{\vec r}_k) \equiv
   \overline{V({\vec r}_1) \cdots V({\vec r}_k)} =\nonumber\\&&
   \int\,
   \left(\prod_{i=1}^S\,\frac{\d{\vec s}_i}{2\pi N}\right)
   \,\prod_{j=1}^k\,\sum_{i_j = 1}^S\,
   W({\vec r}_1 -{\vec s}_{i_1})\cdots W({\vec r}_k-{\vec s}_{i_k}).
\end{eqnarray}

\begin{figure}[ht]
\hbox{
\hspace{0.16\textwidth}\epsfig{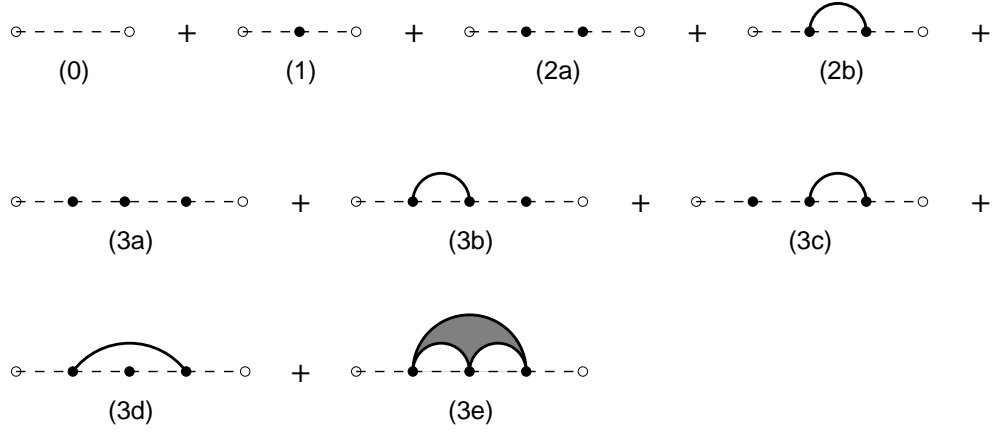}
}
\caption{The diagrams for the averaged Green function, up to $3rd$ order
in the impurity potential. Open dots represent fixed coordinates,
filled dots represent coordinates to be integrated over.
A dashed line (with an implicit direction from left to right)
represent the projection ${\cal P}_{\nu}$. A full line, or filled regions,
represent various correlators of the random potential $V$.}
\label{fig:F0_3}
\end{figure}

\begin{figure}[ht]
\hbox{
\hspace{0.16\textwidth}\epsfig{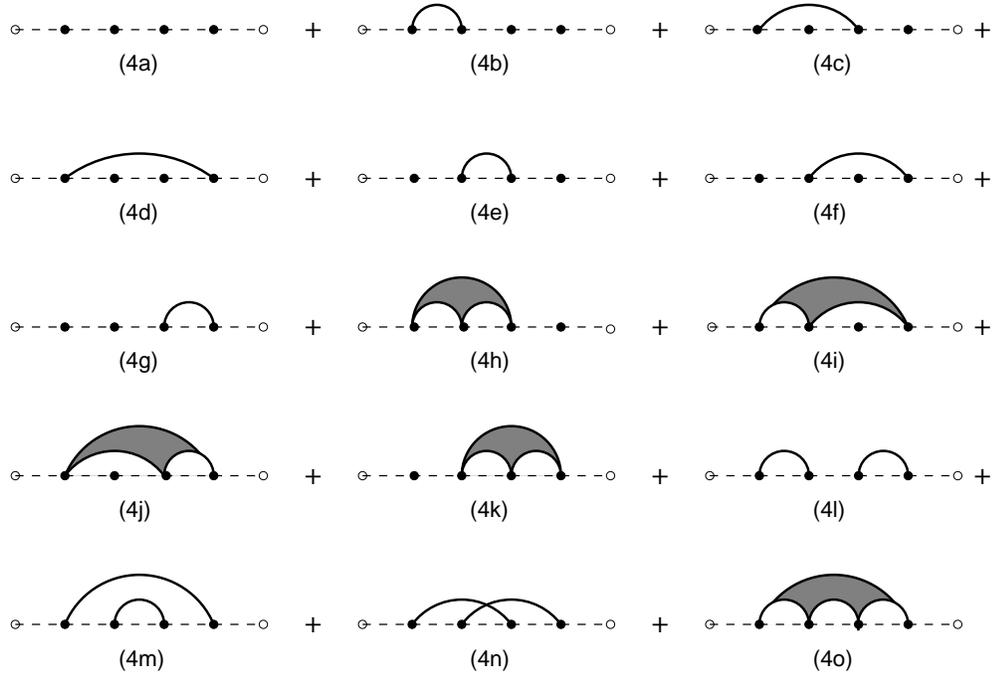}
}
\caption{The $4th$ order diagrams for the averaged Green function.}
\label{fig:F4}
\end{figure}

First consider the situation that all the ${\vec s}_{i_j}$ correspond
to different impurities (as is likely the be the case when all the positions
${\vec r}_j$ are far apart). Then all integrations factorize,
each giving a factor $N^{-1}$. Summing over all possible ways
to pick $k$ different impurities from a
total of $S$, we get the contribution
\begin{equation}
   \label{AllDifferent}
   N^{-k}\left(\begin{array}{c}S\\k\end{array}\right)
   = f(f-\frac{1}{N})\cdots(f-\frac{k-1}{N})
   \equiv f^{(k)} \to f^k \quad\mbox{as $N\to\infty$}
\end{equation}
to $U({\vec r}_1,\ldots,{\vec r}_k)$.
The other extreme is to assume that all the ${\vec s}_{i_j}$ correspond
to same impurity (as is likely to be the case when the
density of impurities is very low, and all the ${\vec r}_i$'s are
close together). Integrating over the impurity position we get a contribution
\begin{equation}
   U_k({\vec r}_1,\ldots,{\vec r}_k) \equiv \frac{f}{k}
   \left(\frac{2}{\varrho-1}\right)^{k-1}
   \exp\left(-\sum_{i,j=1}^k
   \frac{({\vec r}_i-{\vec r}_j)^2}{2k(\varrho-1)}\right)
   \label{U_ks}
\end{equation}
to $U({\vec r}_1,\ldots,{\vec r}_k)$.
The full average must take into account all possible variations between
the above two extremes. The complete expression is (as $N\to\infty$)
\begin{equation}
  \label{FullSeries}
  U({\vec r}_1,\ldots,{\vec r}_k) = \sum_{\boldmath\cal Q}
  U_{a_1}({\vec r}_{.},\ldots,{\vec r}_{.})\cdots
  U_{a_j}({\vec r}_{.},\ldots,{\vec r}_{.}),
\end{equation}
where the sum runs over all partitions $\boldmath\cal Q$ of the set
$\{\vec{r}_1,\ldots,{\vec r}_k\}$ into $j=1,\ldots,k$
nonempty subsets of sizes $a_1,\ldots,a_j$,
and the $U_a$'s are symmetric
functions of the coordinates in each subset. There are
$\sum_{j=1}^k {\cal S}^{(j)}_k$ terms in the sum,
where the ${\cal S}^{(j)}_k$'s are the Stirling numbers
of the second kind\cite{Abramowitz}. Each term correspond
to a Feynman diagram in the perturbation expansion. In our
computer algebra program each $k$'th order Feynman diagram
is represented by a partition of the set $\{1,\ldots,k\}$.
By generating all partitions of this set we obtain all 
$k$'th order diagrams. They are all topologically distinct,
and all occur with combinatorial factor unity.

The 9 diagrams for the $0$'th through $3$'rd order of expansion
are shown in figure~\ref{fig:F0_3},
and the 15 diagrams of $4$'th order are
shown in figure~\ref{fig:F4}.
Here open dots represent fixed coordinates,
and filled dots represent coordinates to be integrated over.
A dashed line (with an implicit direction from left to right)
represent the projection ${\cal P}_{\nu}$. The symmetric, translation
invariant functions $U_q({\vec r}_{1},\ldots,{\vec r}_{q})$ are
represented by filled regions connected to $q$ filled dots.
For $q=1$ they reduce to isolated filled dots, to each of which
there is associated a factor $f$.

According to (\ref{U_ks}-\ref{FullSeries}),
a partition ${\boldmath\cal Q}$ into $j$
nonempty subsets gives a contribution which is proportional to
$f^{j}$. This is fully correct only in the limit $N\to\infty$.
A finite size correction is obtained by making the replacement
$f^j \to f^{(j)}$, cf.\ equation (\ref{AllDifferent}).

The evaluation of the individual diagrams amounts to computing
integrals like
\[
   \int \prod_{i=1}^k \d{\vec \rho}_i\;
   {\cal P}_{\nu}({\vec r},{\vec \rho}_1)
   {\cal P}_{\nu}({\vec \rho}_1,{\vec \rho}_2)
   \cdots
   {\cal P}_{\nu}({\vec \rho}_k,{\vec \rho})\;
   U_{a_1}({\vec \rho}_{.},\ldots,{\vec \rho}_{.})\cdots
   U_{a_j}({\vec \rho}_{.},\ldots,{\vec \rho}_{.})
\]
For the lowest Landau level, $\nu=0$,
this is just a sum of Gaussian integrals.
In the higher levels it becomes a sum of
Gaussians multiplied by polynomials.

Since the averaging process restores translation
invariance, each diagram must be a translation invariant
(with respect to the magnetic translation group)
expression constructed only out of states in the
$\nu$'th Landau level. This forces it
to be proportional to ${\cal P}_{\nu}({\vec r},{\vec\rho})$.
To show this more explicitly, we start with the most
general expansion of an averaged diagram,
\[
    \overline{G_{\boldmath{\cal Q}}({\vec r},{\vec\rho})} =
    \int \d p\,\d q\; C(p,q)\,\psi_p({\vec r})\,\psi_q({\vec \rho})^{*},
\]
where $\{ \psi_p \}$ form an orthonormal basis for the $\nu$'th Landau level.

\begin{figure}[ht]
\hbox{
\hspace{0.16\textwidth}\epsfig{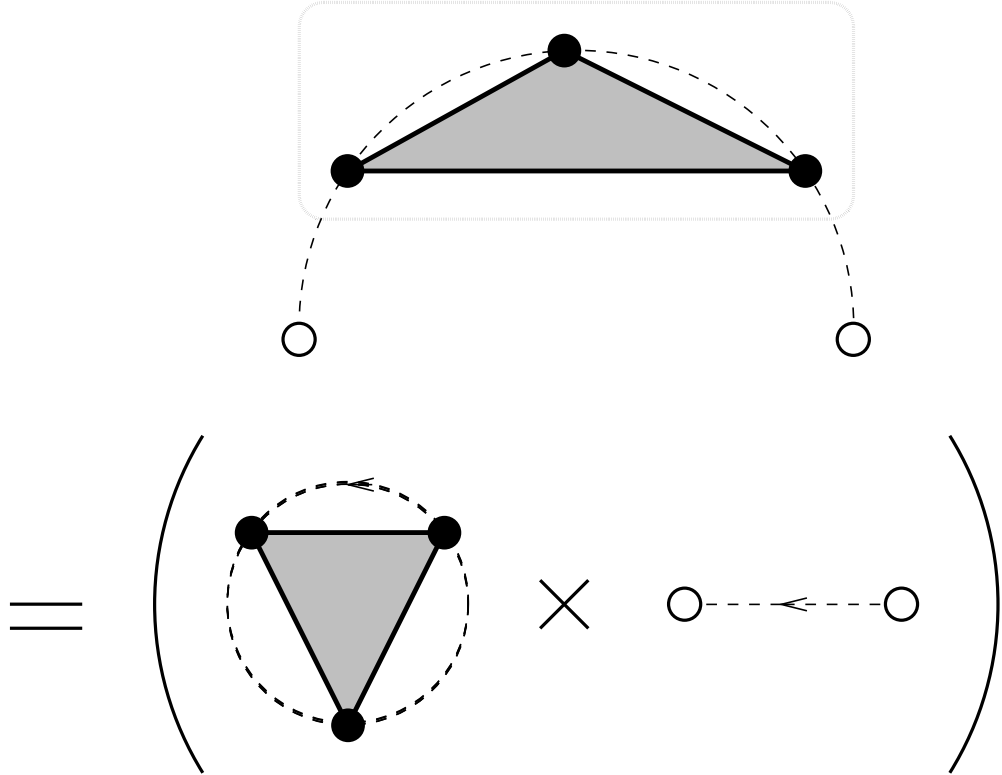}
}
\caption{Due to translation invariance, every diagram contributing to the
averaged Green function becomes proportional to the projection onto the given
Landau level.}
\label{fig:Fbox}
\end{figure}

In the Landau gauge, $A^{y}=0$, we may choose this basis such that the
magnetic translation group acts as
\[
    t(a\hat{x}): \psi_k \rightarrow \mbox{e}^{ika'}\,\psi_k,\qquad
    t(b\hat{y}): \psi_k \rightarrow \psi_{k+b'},
\]
where $a'$ is proportional to $a$, and $b'$ is proportional to $b$.
The requirement that
\[
    \overline{\left[ t(b{\hat y}) G_{\boldmath{\cal Q}}\right]({\vec r},{\vec\rho})}
    = \overline{G_{\boldmath{\cal Q}}({\vec r},{\vec\rho})}
\]
implies that $C(p,q)=C(p-q)$, and the requirement that
\[
    \overline{\left[ t(a{\hat x}) G_{\boldmath{\cal Q}}\right]({\vec r},{\vec\rho})}
    = \overline{G_{\boldmath{\cal Q}}({\vec r},{\vec\rho})}
\]
implies that $C(p-q)=C_0\,\delta(p-q)$, with $C_0$ a constant.
Thus, $\overline{G_{\boldmath{\cal Q}}({\vec r},{\vec\rho})}
\propto {\cal P}_{\nu}({\vec r},{\vec\rho})$.
A graphical representation of this relation is illustrated in
figure~\ref{fig:Fbox} (except that the split-off ``vacuum diagram'' should not be
counted with its usual combinatorial factor).
As a control of our computer algebra program,
we have verified this relation explicitly for a
large number of cases.
Using it reduces the evaluation of $k$'th order diagrams to
calculating determinants of $k\times k$ matrices
(when employing complex coordinates in the lowest Landau level).

Comparing the expansion
\[
  \overline{G({\vec r},{\vec\rho};t)}=
  {\cal P}_{\nu}({\vec r},{\vec\rho})\,\sum_{k=0}^{\infty}\,\frac{(-t)^k}{k!}\,G_k
  \equiv {\cal P}_{\nu}({\vec r},{\vec\rho})\, G(t)
\]
with the general expansion in energy eigenfunctions
\[
  G({\vec r},{\vec\rho};t) = \sum_{\alpha} 
  \psi_{\alpha}({\vec r})\,\psi_{\alpha}^{*}(\vec\rho)\,
  \exp\left(-t E_{\alpha}\right),
\]
we obtain after setting $\vec\rho=\vec r$ and integrating
\begin{equation}
  N\,G(t) = \overline{\sum_{\alpha}\,
  \exp\left(-t E_{\alpha}\right)}
  \equiv N\,\int \d E\,D(E)\,\exp\left(-t E\right),
\end{equation}
where $D(E)$ is the averaged density of states (normalized
to $\int \d E D(E) = 1$.
I.e., $G(t)$ is simply the Laplace transform of $D(E)$, and
the information obtained
from the $k$'th perturbative order is exactly the $k$'th moment of
the averaged density of states,
\begin{equation}
  G_k = \int \d E D(E)\,E^k.
\end{equation}


%% file: moments.tex
\section{Counting and resumming graphs \label{sec:graphcounting}}

We define
\begin{equation}
  {\tilde G}(u) \equiv - \int_0^{\infty} \d t\; \mbox{e}^{t/u}\,G(t) =
  \int \d E\;D(E)\,\frac{u}{1-E u} =
  \sum_{k=0}^\infty G_k\,u^{k+1},
\end{equation}
where the series should be interpreted in an asymptotic sense.
The integral converges for $u<0$, and ${\tilde G}(u)$
can be extended by analytic continuation.
As noted, the graphical expansion illustrated by
figures~\ref{fig:F0_3}--\ref{fig:F4} amounts
to expanding ${\tilde G}$ into a formal power series in $u$.
There are various ways to resum graphical expansions of this type.
The simplest and most useful one is to sum up sequences of ${\cal P}_{\nu}$'s
connected by isolated dots (i.e., factors of $U_1$). This amounts to
making the replacement ${\cal P}_{\nu} \to {\cal P}_{\nu} \,\frac{1}{1-f u}$,
and summing over graphs containing no isolated dots.
The latter sum defines
a function ${\hat G}(u)$, related to ${\tilde G}$ by
\begin{equation}
  \label{SimplerExpansion}
  {\tilde G}(u) = {\hat G}(\frac{u}{1-f u}).
\end{equation}
The $k$'th order graphs for
${\hat G}(u)\sim\sum_{k=0}^\infty {\hat G}_k\,u^{k+1}$
are in 1-1 correspondence
with the partitions of the set $\{ 1,\ldots,k \}$ into subsets
of sizes $a_j \ge 2$. The number of such partitions
is significantly smaller than the total number. For instance,
inspection of figures~\ref{fig:F0_3}--\ref{fig:F4} reveal that only the graphs
$(0)$, $(1)$, $(2b)$, $(3e)$, and $(4\ell)$-$(4o)$ contribute to ${\hat G}$
up to the $4$'th order. The total numbers of graphs which contribute to
${\tilde G}(u)$ ($G_k$) and ${\hat G}(u)$ (${\hat G}_k$)
are listed in \tref{tab:counts}.

\begin{table}
\caption{The number of graphs contributing to various functions.
$G_k$: The total number of graphs contributing to the Green function.
${\hat G}_k$: The number of graphs not containing isolated dots.
${\hat \Sigma}_k$: The number of 1-particle irreducible graphs not
containing isolated dots.
$S_k$: The number of skeleton graphs.
$C_k$: The number of irreducible terms in the answer.
}
\begin{indented}
\item[]\begin{tabular}{@{}rrrrrr}
\br
$k$&$G_k$&${\hat G}_k$&$\hat\Sigma_k$&$S_k$&$C_k$\\
\mr

1&            1&          0&          0&         1&    1\\  \ms
2&            2&          1&          1&         1&    1\\  \ms
3&            5&          1&          1&         1&    1\\  \ms
4&           15&          4&          3&         2&    2\\  \ms
5&           52&         11&          9&         6&    2\\  \ms
6&          203&         41&         33&        21&    7\\  \ms
7&          877&        162&        135&        85&   10\\  \ms
8&         4$\,$140&        715&        609&       385&   36\\  \ms
9&        21$\,$147&       3$\,$425&       2$\,$985&      1$\,$907&   85\\  \ms
10&      115$\,$975&      17$\,$722&      15$\,$747&     10$\,$205&  319\\  \ms
11&      678$\,$570&      98$\,$253&      88$\,$761&     58$\,$455& 1$\,$113\\  \ms
12&     4$\,$213$\,$597&     580$\,$317&     531$\,$561&    355$\,$884& 5$\,$088\\  \ms
13&    27$\,$644$\,$437&    3$\,$633$\,$280&    3$\,$366$\,$567&   2$\,$290$\,$536&     \\  \ms
14&   190$\,$899$\,$322&   24$\,$011$\,$157&   22$\,$462$\,$017&  15$\,$518$\,$391&     \\  \ms
15&  1$\,$382$\,$958$\,$545&  166$\,$888$\,$165&  157$\,$363$\,$329& 110$\,$283$\,$179&     \\  \ms
16& 10$\,$480$\,$142$\,$147& 1$\,$216$\,$070$\,$389& 1$\,$154$\,$257$\,$683& 819$\,$675$\,$482&     \\  \ms
\br
\end{tabular}
\end{indented}
\label{tab:counts}
\end{table}

Another standard resummation is to define $\Sigma(u)$ through the relation
\begin{equation}
   {\tilde G}(u) = \frac{u}{1-\Sigma(u)},
\end{equation}
where only 1-particle irreducible graphs contribute to $\Sigma$. We
may combine this with the resummation of isolated dots above, to
obtain
\begin{equation}
   \Sigma(u) = f u +(1-f u){\hat\Sigma}(\frac{u}{1-f u}).
\end{equation}
Here ${\hat\Sigma}$ is the sum of all 1-particle irreducible graphs
which do not contain isolated dots. The numbers of such graphs are also
listed in \tref{tab:counts}. As can be seen, there is some reduction in the number
of graphs to be calculated,
but not a very significant one. Of the
graphs in figures~\ref{fig:F0_3}--\ref{fig:F4},
the only additional saving is the elimination of
graph $(4\ell)$.

A further reduction in the number of contributing graphs is
obtained by first summing the up the {\em skeleton diagrams} $S(u)$.
This is the set of graphs such that their corresponding ``vacuum
diagrams'' (cf.\ figure~\ref{fig:Fbox}) are 1-particle irreducible.
This function is related to $\Sigma$ by
\begin{equation}
  \label{SkeletonDiagrams}
  \Sigma(u) =\left[1-\Sigma(u)\right]\,S(\frac{u}{1-\Sigma(u)}).
\end{equation}
(The function ${\hat S}(u)=S(u)-f u$ is related to
${\hat\Sigma}(u)$ in the same way.)
The numbers of graphs contributing to $S$ are also
listed in \tref{tab:counts}. 
Expanding $S$ to second order in its argument, and solving the
resulting 2nd degree algebraic equation for $\Sigma$ (or, equivalently, $\tilde G$),
constitutes to the so-called self consistent Born approximation (SCBA).
Of the graphs in figures~\ref{fig:F0_3}--\ref{fig:F4}, the only additional saving
when restricting to skeleton diagrams is the elimination of graph $(4m)$.

One way to employ the resummation methods above is to generate a
smaller set of diagrams (e.g., for ${\hat G}(u)$, ${\hat\Sigma}(u)$, or $S(u)$)
to a given order, and then find the full perturbation expansion to the same
order by use of the algebraic relations above. However, the benefits of going
beyond ${\hat G}(u)$ are marginal, since the reduction in number
of graphs is rather small while more computer
time is needed to classify graphs.
We have in our computation evaluated
all graphs contributing to ${\hat G}(u)$.
As a check on the computer algebra we
have in addition summed the subsets of these graphs contributing to
${\hat\Sigma}(u)$ and $S(u)$, and verified that they reproduce
the same end result for $G(u)$.
The simplest check of the program is to verify that it actually
generates the number of graphs listed in \tref{tab:counts}, since these numbers
are computed in an entirely independent way (starting from the
Stirling numbers).

Another way to use the resummation methods above is to combine a finite
perturbation expansion from $\hat\Sigma(u)$ or $S(u)$ with the exact
algebraic equations, thereby generating some approximate, but infinite,
series for $G(u)$. Considering the numbers in \tref{tab:counts}, we note that
the resummation of isolated dots captures a large percentage of the
total number of graphs. However, the physical effect of this resummation
is rather uninteresting---it merely corresponds to an overall shift
in the energies. Of the remaining graphs we note that the majority of
them are complicated ones (i.e., skeleton graphs) which cannot be
generated by resumming lower order terms. Thus, the resummations above
underestimate the $k$'th perturbative order by increasingly large
amounts as $k$ increases. In the absence of any physical reasons for selecting a
particular class of diagrams, we believe it is much more sensible to
extrapolate the perturbation series in a statistical sense, by
considering how the total number of graphs, and their average value,
varies with the perturbative order $k$.

\section{Calculated moments \label{sec:moments}}

We have calculated the perturbation series up to the $12$'th perturbative
order, for general values of impurity density $f$ and potential range
$\varrho$. Each graph contributes a rational function in $\varrho$, multiplied
by some power of $f$. The general answer rapidly becomes too
complicated to present here. We present the result in terms of the moments $G_k$
of the density of states. The general expressions for
the first five moments are listed in \tref{tab:xi}.
The full result for the case of $\delta$-function impurities, $\varrho=1$,
is listed in \tref{tab:xi0}. We list results for some additional
values of $\varrho$ (in floating point form) 
in \ref{app:CumTabs}.

\begin{table}
\caption{The first five energy moments for general $\varrho$,
in the lowest Landau level.}
\begin{indented}
\item[]\begin{tabular}{@{}rl}
\br
$k$&$G_k$\\
\mr
1 & $f$  \\  \ms
2 & ${\frac {1}{\varrho}}f+{f}^{2}$  \\    \ms
3 & ${\frac {4}{3\,{\varrho}^{2}+1}}f+{\frac {3}{\varrho}}{f}^{2}+{f}^{3}$  \\    \ms
4 & ${\frac {2}{\varrho\,\left ({\varrho}^{2}+1\right )}}f+{\frac {25\,{\varrho}^{4
       }+25\,{\varrho}^{2}+2}{{\varrho}^{2}\left ({\varrho}^{2}+1\right )\left (3\,{
       \varrho}^{2}+1\right )}}{f}^{2}+{\frac {6}{\varrho}}{f}^{3}+{f}^{4} $ \\    \ms
5 &  ${\frac {16}{5\,{\varrho}^{4}+10\,{\varrho}^{2}+1}}f+{\frac {210\,{\varrho}^{4}
       +420\,{\varrho}^{2}+170}{\varrho\,\left (3\,{\varrho}^{2}+1\right )\left (3\,{
       \varrho}^{2}+5\right )\left ({\varrho}^{2}+1\right )}}{f}^{2}+{\frac {85\,{
       \varrho}^{4}+85\,{\varrho}^{2}+10}{{\varrho}^{2}\left ({\varrho}^{2}+1\right )
       \left (3\,{\varrho}^{2}+1\right )}}{f}^{3}+{\frac {10}{\varrho}}{f}^{4}+{f}^{5} $ \\   \ms
\vdots & \\ \ms
$k$ & $\frac{2^k}{(\varrho+1)^k-(\varrho-1)^k}f + \cdots + 
{k\choose3}\frac{(9k-11)(\varrho^4+\varrho^2)+(2k-6)}{
4\varrho^2(\varrho^2+1)(3\varrho^2+1)}f^{k-2}+
\frac{k(k-1)}{2\varrho}f^{k-1}+f^k$\\ \ms 
\br
\end{tabular}
\end{indented}
\label{tab:xi}
\end{table}

\begin{table}
\caption{The twelve first energy moments for $\delta$-function impurities
($\varrho=1$).}
\begin{indented}
\item[]\begin{tabular}{@{}rl}
\br
$k$ & $G_k$ \\
\mr
1 & $f $ \\   \ms
2 & $f+{f}^{2}$  \\   \ms
3 & $f+3\,{f}^{2}+{f}^{3} $ \\   \ms
4 & $f+{\frac {13}{2}}{f}^{2}+6\,{f}^{3}+{f}^{4} $ \\   \ms
5 & $f+{\frac {25}{2}}{f}^{2}+{\frac {45}{2}}{f}^{3}+10\,{f}^{4}+{f}^{5}$  \\   \ms
6 & $f+{\frac {137}{6}}{f}^{2}+{\frac {277}{4}}{f}^{3}+{\frac {115}{2}}{f}
      ^{4}+15\,{f}^{5}+{f}^{6} $ \\   \ms
7 & $f+{\frac {245}{6}}{f}^{2}+{\frac {2317}{12}}{f}^{3}+{\frac {1029}{4}}
      {f}^{4}+{\frac {245}{2}}{f}^{5}+21\,{f}^{6}+{f}^{7} $ \\   \ms
8 & $f+{\frac {871}{12}}{f}^{2} +{\frac {3067}{6}}{f}^{3} +{\frac {24131}{24}}{f}^{4} 
       +749 {f}^{5} +231 {f}^{6} +28 {f}^{7} + {f}^{8}  $  \\   \ms
9 & $f+{\frac {517}{4}}{f}^{2} +{\frac {47443}{36}}{f}^{3} +{\frac {29091}{8}}{f}^{4} 
       +{\frac {31121}{8}}{f}^{5} +1848 {f}^{6} + 399 {f}^{7} 
       +36 {f}^{8} + {f}^{9}$  \\   \ms
10 & $f+\frac{4629}{20}f^2+\frac{241715}{72}f^3+\frac{452455}{36}f^4+\frac{292901}{16}f^5
+\frac{98261}{8}f^6+\frac{8085}{2}f^7+645 f^8 $\\ \ms
   & $+ 45 f^9 + f^{10}$\\ \ms
11 & $f+\frac{8349}{20}f^2+\frac{3068197}{360}f^3+\frac{3047209}{72}f^4 +
      \frac{11665093}{144}f^5+\frac{1163173}{16}f^6+\frac{267377}{8}f^7$\\ \ms
   &   $+\frac{16137}{2}f^8 +990f^9+55 f^{10} + f^{11}$\\ \ms
12 & $f+\frac{45517}{60}f^2+\frac{5201203}{240}f^3+\frac{303556067}{2160}f^4 +
      \frac{16554385}{48}f^5 + \frac{114943133}{288}f^6$\\ \ms
   & $+ \frac{1940433}{8}f^7 +
      \frac{649231}{8} f^8
      + 14982 f^9 + \frac{2915}{2}f^{10} + 66 f^{11} + f^{12}$\\ \ms
\br
\end{tabular}
\end{indented}
\label{tab:xi0}
\end{table}

It is an amusing exercise to search for general patterns in
\tref{tab:xi} and \tref{tab:xi0}. The coefficients of
the highest powers of $f$ are straightforward to find, since
the series for ${\hat G}(u)$ up to the $2j$'th perturbative order
determines all the terms proportional to $f^k$, $f^{k-1}$,
\ldots, $f^{k-j}$ in the $k$'th perturbative order.
The lowest powers of $f$ are of more physical interest. Consider
first the case of $\delta$-function impurities. The general pattern
of the $f^2$-terms looks a bit complicated, but it is easy to verify
that it fits the formula
\begin{equation}
   G_k = f + f^2\,\sum_{p=1}^{\lfloor k/2\rfloor}\,\frac{1}{p}
   \left(\begin{array}{c}k\\2p\end{array}\right) + \Or(f^3).
\end{equation}
These moments are reproduced to order $f^2$ by the density of
states
\begin{equation}
   \label{LowDensityDistribution}
   D(E) = (1-f)\,\delta(E)+
   \case{1}{2}\,f^2\,\theta(E)\,\theta(2-E)\,\vert E -1\vert^{f-1}.
\end{equation}
The physics behind this expression is as follows:
\begin{enumerate}

\item
Consider $S=f N < N$ arbitrary placed impurities in a
single Landau level, for which the Hilbert space of wave functions
is $N$-dimensional. The condition that a wave function vanishes on
all impurities imposes $S$ constraints, for which the solution space
is at least $(N-S)$-dimensional. These solutions will have exactly zero
energy, which is also the lowest possible energy. Hence, there must for
all $f<1$ be a $\delta$-function contribution to the density of states,
of strength $(1-f)$.

\item
At very low $f$ the impurities are very far apart. Each of them
will bind one state of energy $E=1$. Hence, they will
contribute a term $f\,\delta(E-1)$ to the density of states.
This is sufficient to reproduce all moments to order $f$. To order
$f^2$ we must consider the mixing between states at different
impurities (to this order only mixing between
two impurities). The energies for two impurities
at distance $r$ are $E_{\pm}=1\pm \mbox{e}^{-r^2/4}$.
Integrating over the distribution of relative distances
we get a contribution to the density of states,
\begin{eqnarray*}
   &&\Delta D(E) \propto \int \d{\vec r}\;
   \left[\delta(E-1-\mbox{e}^{-r^2/4}) +
         \delta(E-1+\mbox{e}^{-r^2/4})
   \right]\\
   &&\propto \int_0^1 \frac{\d x}{x}\, \left[\delta(E-1-x)+
   \delta(E-1+x)\right]\\
   && \propto\;
   \theta(E)\,\theta(2-E)\,\vert E -1 \vert^{-1}.
\end{eqnarray*}
This distribution is not normalizable, but has finite moments
$\langle (E-1)^k\rangle$. The lack of normalization is due to
the fact that additional impurities must be taken into account at
very large separations $r$ (of order $f^{-1/2}$). A simple way to model
this is by introducing a  distribution $dp(r)=\mbox{e}^{-kfr^2}kf dr^2$
for the relative pair distance (the total density of pairs
being $\frac{1}{2}f$). The requirement that all moments be reproduced
to order $f^2$ fixes $k=\frac{1}{4}$. This in turn leads to
(\ref{LowDensityDistribution}).

\item
If one repeat the previous argument with the finite size correction,
$f^2\to f^{(2)} = f(f-N^{-1})$, one is lead to the conclusion
that the exponent in (\ref{LowDensityDistribution}) should
also undergo a finite size correction,
\[
   {\vert E-1 \vert}^{f-1} \to {\vert E-1 \vert}^{f-1-1/N}.
\]

\end{enumerate}

It is also easy to understand the physical origin of the
low-$f$ behaviour of the general moments,
\begin{equation}
    \label{GeneralLowfMoments}
    G_k = \frac{2^k}{(\varrho+1)^k-(\varrho-1)^k}\,f + \Or(f^2).
\end{equation}
This is related to the behaviour of electrons near the single impurity
potential (\ref{OneImpurityPotential}). With the impurity
at the origin we find eigenstates (in the symmetric gauge, and cylinder
coordinates)
\[
   \psi_\ell (\vr)= (2 \pi \, 2^\ell \, \ell!)^{-1/2}\;
   r^\ell \, \e^{-\i\ell\varphi} \, \e^{-r^2/4},
\]
and corresponding energies (cf.\ Appendix~A)
\begin{equation}
   E_\ell = \left(\frac{2}{\varrho+1}\right)
   \left(\frac{\varrho-1}{\varrho+1}\right)^{\ell},\qquad \ell=0,1,2,\ldots.
\end{equation}
With a small fraction $f$ of such impurities the contribution to the density
of states becomes
\begin{equation}
   D(E) = f\,\sum_\ell \delta(E-E_\ell).
\end{equation}
The sum must be cut off when the extension
$r_\ell$ of the wave function becomes of
order $f^{-1/2}$, i.e.\ for $\ell$ of order $f^{-1}$. This is important for
obtaining a normalized $D$, but can be ignored when calculating
its moments. Thus we get to order $f$
\[
   G_k = f\;\sum_{\ell=0}^{\infty}\, E_\ell^k =
   f\;\left(\frac{2}{\varrho+1}\right)^k\,
         \left[1-\left(\frac{\varrho-1}{\varrho+1}\right)^k\right]^{-1}.
\]

\subsection{The cumulant expansion}

The information in \tref{tab:xi} and \tref{tab:xi0} can be compressed somewhat
by rewriting it in terms of cumulants. The moments
$\langle E^k \rangle$ are related to the Laplace transform of the
density of states,
\begin{equation}
   G(t) \equiv \int_0^{\infty} \d E \,D(E)\,
   \e^{-t E} = \sum_{k=0}^{\infty} \frac{(-t)^k}{k!}\,G_k.
   \label{LaplaceTransform}
\end{equation}
By rewriting
\begin{equation}
   G(t) = \exp\left(\chi(t)\right) =
   \exp\left(\sum_{k=1}^{\infty} t^k\,\chi_k\right)
   \label{CumulantExpansion}
\end{equation}
we obtain the cumulant expansion. No information is lost when going
between the set of cumulants $\{ \chi_k \vert k=1,\ldots,m\}$
and the set of moments $\{G_k \vert k=1,\ldots,m \}$. 

\begin{table}
\caption{The first five cumulants for general $\varrho$,
in the lowest Landau level.}
\begin{indented}
\item[]\begin{tabular}{@{}rl}
\br
$k$ & $(-1)^k \, k! \, \chi_k$ \\
\mr
1 & $f $ \\  \ms
2 & ${\frac {1}{\varrho}}f $ \\  \ms
3 & ${\frac {4}{3\,{\varrho}^{2}+1}}f $ \\  \ms
4 & ${\frac {2}{\varrho\,\left ({\varrho}^{2}+1\right )}}f-{\frac {1}{\left ({
       \varrho}^{2}+1\right ){\varrho}^{2}}}{f}^{2} $ \\  \ms
5 & ${\frac {16}{5\,{\varrho}^{4}+10\,{\varrho}^{2}+1}}f-{\frac {80}{\left (3\,{
        \varrho}^{2}+5\right )\varrho\,\left (3\,{\varrho}^{2}+1\right )}}{f}^{2}$ \\  \ms
$\vdots$ & \\ \ms
$k$ & $ \frac{2^k}{(\varrho+1)^k-(\varrho-1)^{k}}f+\cdots$ \\ \ms
\br
\end{tabular}
\end{indented}
\label{tab:chi}
\end{table}

\begin{table}
\caption{The 12 first cumulants for $\delta$-function impurities,
in the lowest Landau level.}
\begin{indented}
\item[]\begin{tabular}{@{}rl}
\br
$k$& $(-1)^k \, k! \, \chi_k$ \\
\mr
1 & $f $ \\  \ms
2 & $f $ \\  \ms
3 & $f $ \\  \ms
4 & $f-{\frac {1}{2}}{f}^{2} $ \\  \ms
5 & $f-{\frac {5}{2}}{f}^{2} $ \\  \ms
6 & $f-{\frac {49}{6}}{f}^{2}+{\frac {7}{4}}{f}^{3} $ \\  \ms
7 & $f-{\frac {133}{6}}{f}^{2}+{\frac {77}{4}}{f}^{3} $ \\  \ms
8 & $f-{\frac {653}{12}} {f}^{2}+{\frac {757}{6}} {f}^{3}-{\frac {109}{8}} {f}^{4} $ \\  \ms
9 & $f-{\frac {503}{4}} {f}^{2}+{\frac {11603}{18}} {f}^{3}-{\frac {2085}{8}} {f}^{4} $ \\  \ms
10& $f-\frac{5591}{20}f^2+\frac{204725}{72}f^3-\frac{17065}{6}f^4+\frac{2971}{16}f^5$ \\ \ms
11& $f-\frac{12111}{20}f^2+\frac{820787}{72}f^3-\frac{841489}{36}f^4+\frac{87197}{16}f^5$ \\ \ms
12& $f-\frac{77303}{60}f^2+\frac{10270399}{240}f^3-\frac{69677951}{432}f^4+\frac{87197}{16}f^5-
     \frac{124513}{32}f^6$ \\ \ms
\br
\end{tabular}
\end{indented}
\label{tab:chi0}
\end{table}

We show the five first cumulants for general values of $f$ and $\varrho$
in \tref{tab:chi}, and the twelve first cumulants for $\delta$-function
potentials ($\varrho=1$) in \tref{tab:chi0}.
Note that a factor $(-1)^k/k!$ has been split off in both tables.
The cumulant expansion may also be viewed as yet another way of resumming a
perturbation expansion. We return to this below.


%% file: brezin.tex
\section{Comparison with the exact results of
Br\'{e}zin et.\ al.\ and Wegner\label{sec:brezin}}

For the case of $\delta$-function impurities, $\varrho=1$,
exact formulas for the density of states have been found
by Br\'{e}zin {\em et. al.}\cite{Brezin}, and by Wegner\cite{Wegner}.
The result by Wegner correspond to taking the additional limit of $f\to\infty$.
Their expressions for the density of states are highly non-trivial,
and they provide very useful checks on our results and methodology.

\subsection{The result of Br\'{e}zin {\em et.\ al.}}

\begin{figure}[ht]
\hbox{
\hspace{0.16\textwidth}\epsfig{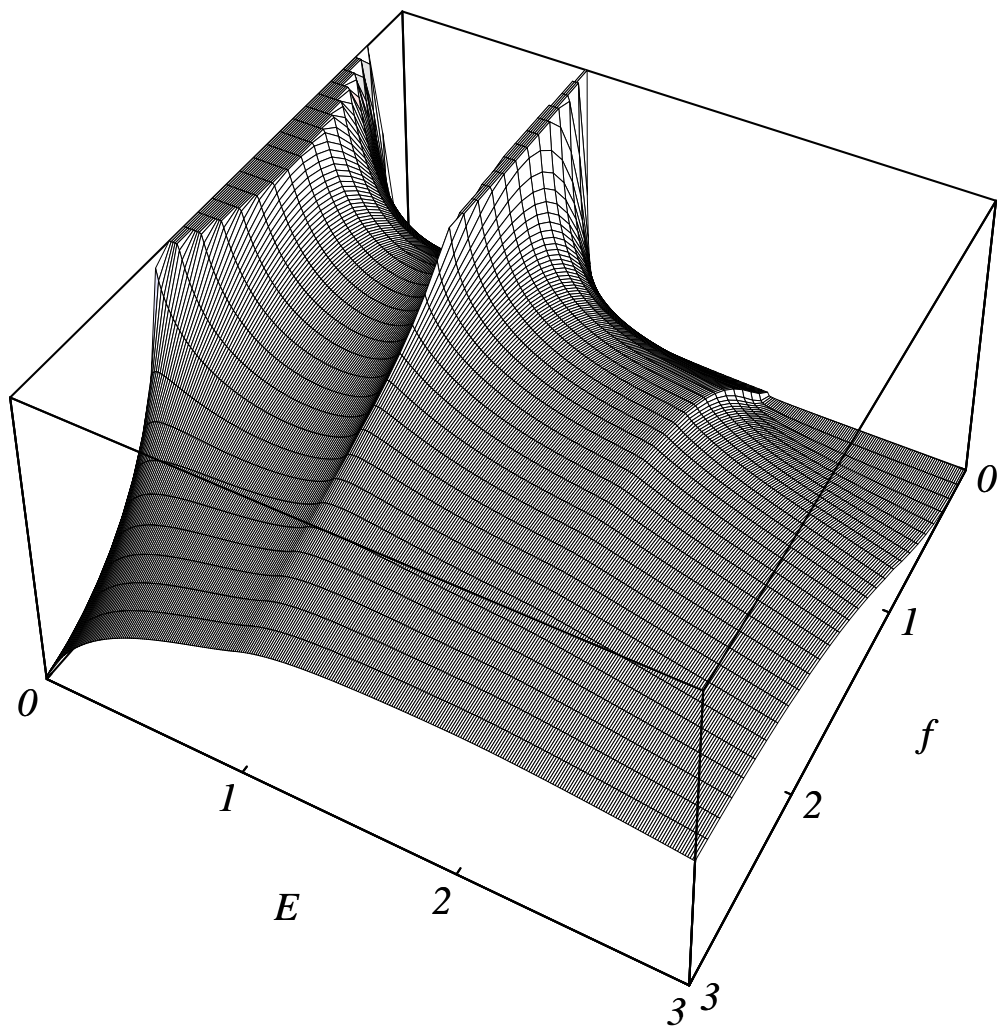}
}
\caption{A $3D$-plot of the density of states found by Br\'{e}zin {\em et.\ al.},
as function of energy $E$ and impurity density $f$. In addition to the
plotted density there is a contribution
$(1-f)\,\delta(E)$ when $f<1$.}
\label{fig:Brezinplot1}
\end{figure}

The results found in reference \cite{Brezin} specializes for our model
(with $\varrho=1$) to the integral
\bea
  D_{\B} (E)=\frac{1}{\pi} \, \Im \frac{\partial}{\partial E} \ln 
       \int_{0}^{\infty} \d t \, \exp[\, \i E t -f\, I(t)\,]\, ,\qquad {\rm where} \nonumber \\
  I(t)=\int_{0}^{t} \frac{\d \beta}{\beta} \left( 1-\e^{-\i \beta} \right)
          =\i t + {\tsty \frac{1}{4}}\, t^2 - {\tsty \frac{\i}{18}}\, t^4 - \cdots.
  \label{eq:rhoBfull}
\eea
It is a rather challenging task to evaluate this expression numerically.
As was pointed out in \cite{Brezin}, $D_{\B}(E)$ has a lot of interesting
structure, in particular for low density of impurities $f$:

\begin{enumerate}

\item For $f<1$ there is a $\delta$-function contribution to $D_{\B}(E)$,
\begin{equation}
  D_{\B}(E) = (1-f)\,\delta(E) + \cdots.
\end{equation}
As explained before (cf.\ the discussion around
equation (\ref{LowDensityDistribution})), this is due to the fact that
we may arrange for a fraction $1-f$ of all wavefunctions to vanish on
all impurities. This phenomenon was already pointed out by
Ando\cite{Ando}.

\item
As $E\to 0^+$ one finds the behaviour
\begin{equation}
  D_{\B}(E) \sim \left\{
  \begin{array}{ll}
     E^{-f} & 0<f<1,\\
     E^{-1}/\log^{2}(E) & f=1,\\
     E^{f-2} & 1<f.
  \end{array}
  \right.
\end{equation}
The increasingly singular behaviour as $f$ increases from $0$ to $1$ is
due to zero-energy states moving out into the low $E$ region.
As for the large $f$-dependence, a
crude qualitative understanding is obtained by considering an arbitrary placed,
maximally localized state,
\[
  \vert \psi({\vec r})\vert^2 = 
  {\cal N}\,\e^{-({\vec r}-{\vec r}_0)^2/2}.
\]
The probability $\mbox{Prob}(E < E_0)$ that this state has an energy less that $E_0$
is equal to the probability that the distance $r$ from
${\vec r}_0$ to its nearest impurity satisfies
${\cal N}\,\exp(-r^2/2) < E_0$. Now,
\[
   \mbox{Prob}\left(\exp(-r^2/2)
   < E_0/2\pi{\cal N}\right) = \left({E_0}/2\pi{\cal N}\right)^f,
\]
which predicts $D(E)\sim E^{f-1}$. This is an underestimate
(because our assumption that the state was maximally localized,
and arbitrary placed, cannot be expected to hold),
but the dependence on $f$ turns out to be correct.

\item
As $E\to 1$ one finds a diverging density of states
\begin{equation}
  D_{\B}(E) \sim \vert E-1 \vert^{f-1},\quad \mbox{for $0<f<1$,}
\end{equation}
and this singularity continues as a cusp singularity for $1<f<2$. As discussed
before (cf.\ equation (\ref{LowDensityDistribution})),
this singularity originates in states localized on top of single
impurities. The interaction with nearby impurities leads to a level
broadening around $E=1$.

\item
As $E\to 2$ there is a cusp singularity,
\begin{equation}
  \frac{d}{dE}D_{\B}(E) \sim  \vert E-2 \vert^{f-1},\quad \mbox{for $0<f<1$.} 
\end{equation}
This singularity is due to the fact that a state must interact with at least
three impurities to have an energy $E>2$, while two impurities are sufficient
for $E<2$. By a similar reasoning one can also understand why there are (increasingly
weak) singularities in $D_{\B}(E)$ at $E=3,4,\ldots$. 

\item
$D_{\B}(E)$ is a fairly smooth function for $f>2$.

\end{enumerate}

\begin{figure}[ht]
\hbox{
\hspace{0.16\textwidth}\epsfig{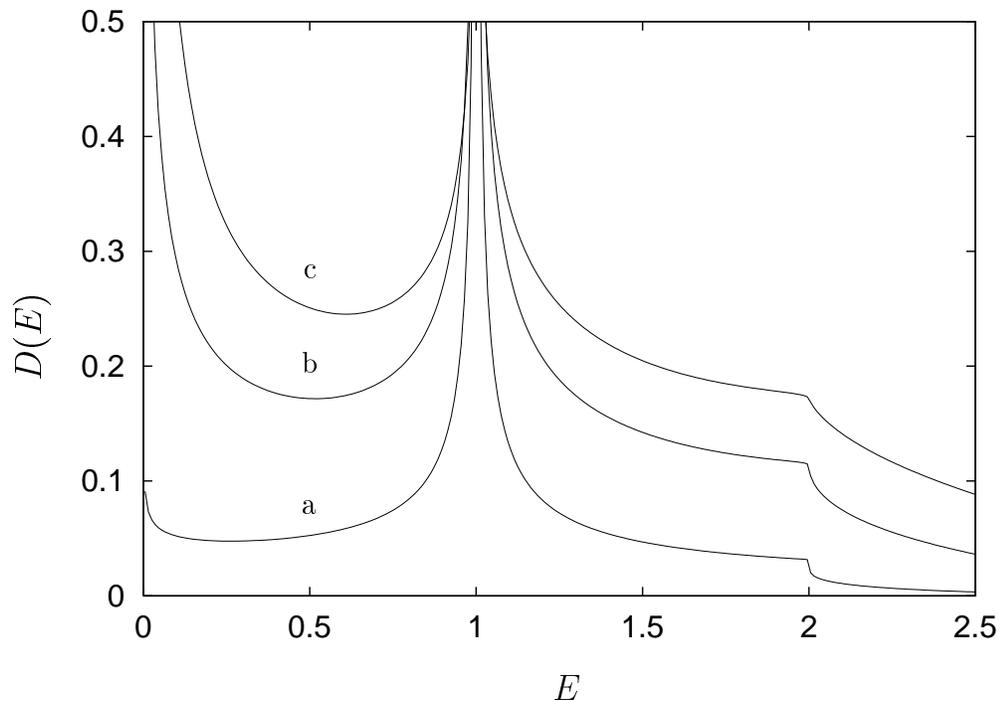}
}
\caption{The density of state for a) $f=0.2$, b) $f=0.5$, and c) $f=0.8$.
Due to a $\delta$-function contribution at $E=0$ the curves are normalized
to $f$ instead of 1.
}
\label{fig:Brezinplot2}
\end{figure}
 
Plots for the integrated density of states, $\int_0^E \d E' D_{\B}(E')$, where given
in \cite{Brezin} and \cite{ItzyksonDrouffe}. The plots of $D_{\B}(E)$ itself look more
interesting; we give examples in figures~\ref{fig:Brezinplot1}-\ref{fig:Brezinplot2}.
We shall return to the a reconstruction of $D_{\B}(E)$ from our calculated
moments, but first consider the limiting case of $f\to\infty$.

\subsection{The limit of high impurity density, and the result of Wegner}

Consider now equation (\ref{eq:rhoBfull}) for very large $f$.
The main contribution to the integral will come from small $t$, hence
we may approximate
\begin{equation}
   I(t) \approx it+\case14 t^2.
\end{equation}
This leads to the expression
\bea
  &&
  \sqrt{f}\,D_{\B}(E) \approx D_{\W}(\epsilon) \equiv
  \frac{1}{\pi} \, \Im \frac{\partial}{\partial \epsilon} 
       \ln \left(1+ \mbox{erf} (\i \epsilon) \right) \nonumber \\
       &&= \frac{2}{\pi^{3/2}} \, \frac{\exp(\epsilon^2)}{1-\mbox{erf}^{\,2} (\i \epsilon)} 
            = \frac{2}{\pi^{3/2}} \, \frac{\exp(\epsilon^2)}{1+\mbox{irf}^{\,2} (\epsilon)},
  \label{eq:rhoB}
\eea
where $\epsilon=(E-f)/\sqrt{f}$,
and $\mbox{irf}(\epsilon)\equiv -\i \,\mbox{erf}(\i \epsilon)$
is a real function of a real argument. This is equivalent to the
result by Wegner\cite{Wegner}.

The Fourier transform of $D_{\W}(\epsilon)$ has the series expansion
\begin{eqnarray} 
  &{\tilde D}_{\W} (\omega) = \int_{-\infty}^{\infty} \d \epsilon \, 
        \e^{-\i \, \omega \, \epsilon} D_{\W} (\epsilon)\nonumber\\
  &= 1 - {\tsty \frac{1}{2!}} \,\langle \epsilon^{2} \rangle\, \omega^{2}
            +{\tsty \frac{1}{4!}} \,\langle \epsilon^{4} \rangle\, \omega^{4}
            -{\tsty \frac{1}{6!}} \,\langle \epsilon^{6} \rangle\, \omega^{6}
            +{\tsty \frac{1}{8!}} \,\langle \epsilon^{8} \rangle\, \omega^{8} -\cdots,
  \label{FTWegner}
\end{eqnarray}
with $\langle \epsilon^{n} \rangle = \int_{-\infty}^{\infty} \d \epsilon \,
D_{\W} (\epsilon) \, \epsilon^{n}$.
One may verify (numerically) that 
\be
\fl
  \langle \epsilon^{2} \rangle = 1,\ 
  \langle \epsilon^{4} \rangle = \case52,\  
  \langle \epsilon^{6} \rangle = {\tsty \frac{37}{4}},\  
  \langle \epsilon^{8} \rangle = {\tsty \frac{353}{8}},\ 
  \langle \epsilon^{10} \rangle = {\tsty \frac{4081}{16}},\ 
  \langle \epsilon^{12} \rangle = {\tsty \frac{55205}{32}}.  
  \label{eq:e28}
\ee
All odd moments vanish because $D_{\W}(\epsilon)$
is an even function of $\epsilon$.

Now turn to the problem of determining $D(E)$ from e.g.\ the cumulants in
table \ref{tab:chi}. Formally this may be done by inverting the
Laplace transform (\ref{LaplaceTransform}-\ref{CumulantExpansion}).
With $t=\i u$ this becomes
\be
   D(E)={\frac {1}{2 \pi}} \int_{-\infty}^{\infty} \d u \, \e^{i(E-f)u}\,
   \exp \sum_{k=2}^\infty (\i u)^k \chi_k.
  \label{eq:rhof}
\ee
To investigate the large-$f$ limit we introduce $\epsilon=(E-f)/\sqrt{f}$,
$\omega=u\sqrt{f}$, and keep only terms which remain
when $f\to\infty$. Thus, in this limit,
\be
   \sqrt{f}\,D(E)  \approx \frac{1}{2\pi} \int_{-\infty}^{\infty} \d\omega\,
   \e^{\i\epsilon\omega}\,{\tilde D}(\omega),
\ee
where we find from table \ref{tab:chi}
\be\fl
   {\tilde D}(\omega) = \exp\left(
   -\case12\, \omega^2
   -{\tsty \frac{1}{2\cdot 4!}}\, \omega^{4} 
   -{\tsty \frac{7}{4\cdot6!}}\, \omega^{6} 
   -{\tsty \frac{109}{8\cdot8!}}\, \omega^{8}
   -{\tsty \frac{2971}{16\cdot10!}}\, \omega^{10}
   -{\tsty \frac{124513}{32\cdot12!}}\, \omega^{12}
   - \cdots 
   \right).
   \label{CumulantExpansion2}
\ee
One may verify that $D_{\W}(\omega)=D(\omega)$ to the order in $\omega$ we have
computed.

\subsection{From cumulant expansion to density of states}

\begin{figure}[ht]
\hbox{
\hspace{0.16\textwidth}\epsfig{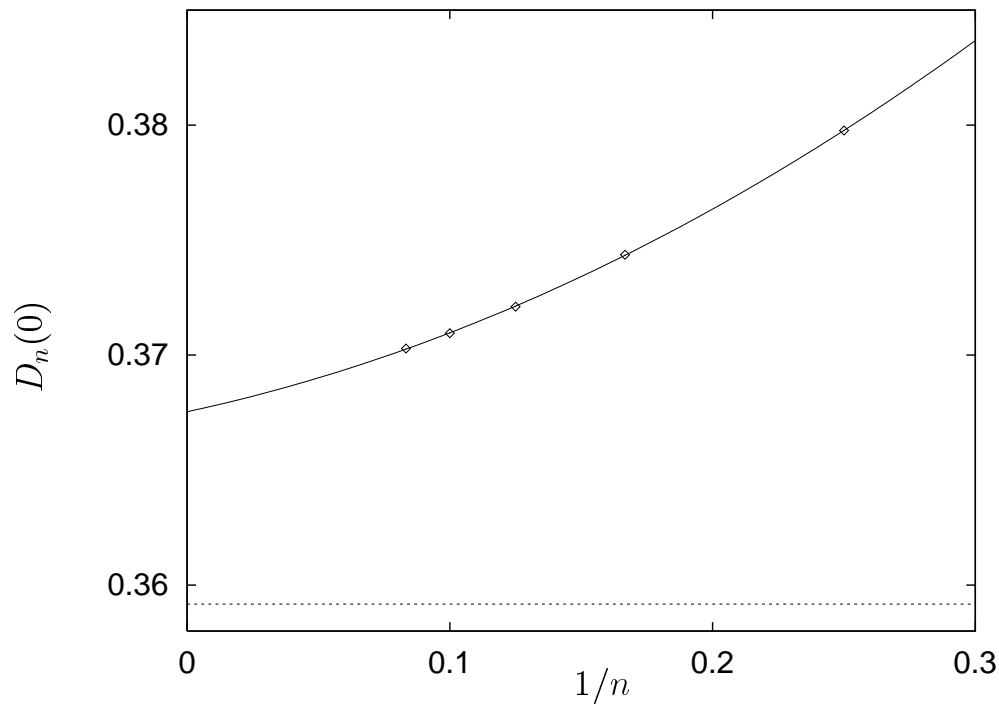}
}
\caption{The approximants $D_n(0)$ as function of $n^{-1}$, together
with a fitting function. It is clear that the approximants converges
to a value which is different from the (indicated) exact result,
$D_{\W}(0)=0.35917\ldots$}
\label{fig:Fconvergence}
\end{figure}

We now construct approximants $D_n(\epsilon)$ to $D(\epsilon)$, using
equation (\ref{eq:rhof}) with the cumulants $\chi_2,\ldots,\chi_n$
included in the sum. The result is surprising and instructive.
To investigate convergence we evaluate the densities at $\epsilon=0$
for increasing $n$. We find
\[ 
   D_n(0) =(0.3788,0.3744,03721,0.3710,0.3703) \mbox{ for $n=(4,6,8,10,12)$.}
\]
This sequence converges nicely with $n$
(see figure~\ref{fig:Fconvergence}),
fitting the formula
\[
    D_n(0) \approx 0.3675+0.02456\,n^{-1}+0.09745\,n^{-2}+\cdots.
\]
However, when extrapolated to $n=\infty$ we obtain $D_{\infty}(0)\approx 0.3675$,
which is about $2.3$ percent higher than the exact result,
\[
  D_{\W}(0)=2/\pi^{3/2}=0.35917\ldots.
\]

\begin{figure}[ht]
\hbox{
\hspace{0.16\textwidth}\epsfig{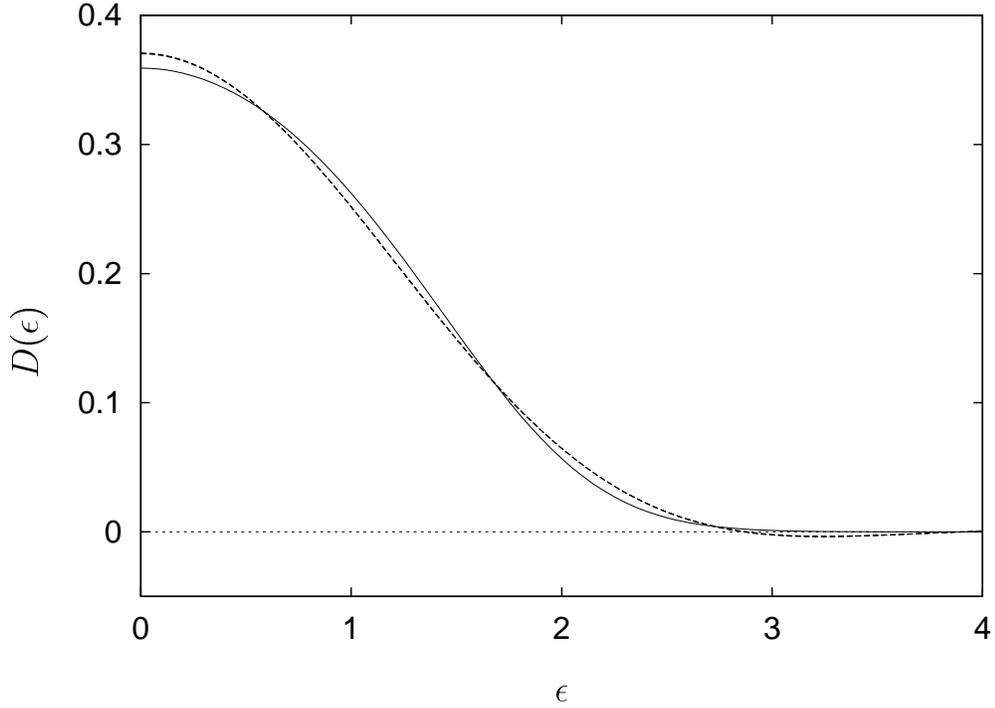}
}
\caption{A comparison of the approximants $D_{10}(\epsilon)$ and
$D_{12}(\epsilon)$ (dashed lines) with the exact result $D_{\W}(\epsilon)$
(fully drawn line).
Since all curves are symmetric with respect to $\epsilon\to-\epsilon$
only positive $\epsilon$ is plotted.
$D_{10}$ and $D_{12}$ are almost indistinguishable, indicating good convergence.
However, the approximants have converged to a result
which is different from the exact one.}
\label{fig:FrhoE}
\end{figure}

The situation is further illustrated in figure~\ref{fig:FrhoE}, where we
compare the functions $D_{10}(\epsilon)$,
$D_{12}(\epsilon)$ and $D_{\W}(\epsilon)$. The curves for
$D_{10}(\epsilon)$ and $D_{12}(\epsilon)$ are almost
indistinguishable. This indicates that the sequence of approximants has
converged. However, the limit function turns out to be {\em negative}
for certain ranges of $\epsilon$. Thus, even without knowledge of the
exact result we could have concluded that something was wrong.

\begin{figure}[ht]
\hbox{
\hspace{0.16\textwidth}\epsfig{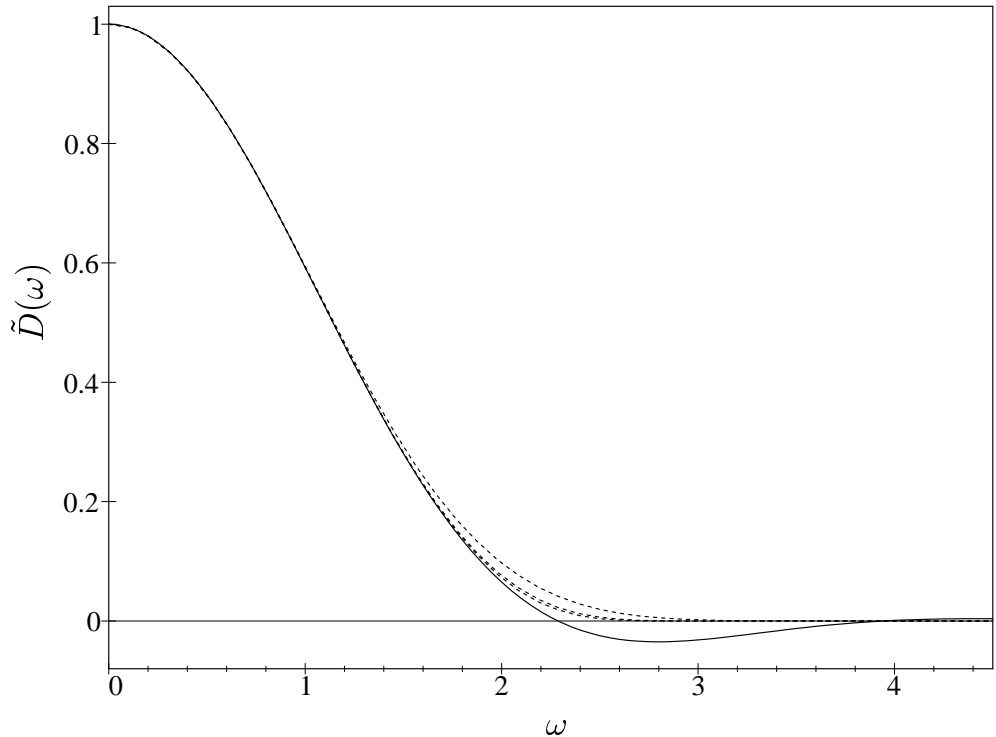}
}
\caption{Fourier transforms of density of states. The exact result
${\tilde D}_{\W}(\omega)$ (fully drawn line), is compared with the
approximants ${\tilde D}_4(\omega)$, ${\tilde D}_8(\omega)$ and
${\tilde D}_{12}(\omega)$ constructed from the cumulant expansion.
The exact function becomes negative in some regions, a property
which cannot be reproduced by the approximants.
}
\label{fig:Flog}
\end{figure}

The origin of the problem can be understood by investigating the
behaviour of ${\tilde D}_{\W}(\omega)$. This is plotted in
figure~\ref{fig:Flog}, together with the approximants
${\tilde D}_4(\omega)$, ${\tilde D}_8(\omega)$ and
${\tilde D}_{12}(\omega)$. We observe that
${\tilde D}_{\W}(\omega)$ becomes negative for some
$\vert\omega\vert>\omega_0\approx 2.3$.
Since all the $\chi_k$ are real, the cumulant expansion
can never reproduce such a behaviour. In fact, all terms in the
exponent in equation (\ref{CumulantExpansion2})
seem to have the same (negative) sign.
Thus, we expect the series in the exponent to converge to
$\log {\tilde D}_{\W}(\omega)$ for $\vert\omega\vert<\omega_0$,
and diverge to $-\infty$ for $\vert\omega\vert > \omega_0$.
Thus, our approximants ${\tilde D}_n(\omega)$ will converge to
the (wrong) limit
\be
       {\tilde D}_{\infty}(\omega) = \left\{\begin{array}{rl}
       {\tilde D}_{\W}(\omega), & \mbox{for $\vert\omega\vert<\omega_0$,}\\
       0,                       & \mbox{for $\vert\omega\vert>\omega_0$.}
       \end{array}\right.
\ee
The curves in figure~\ref{fig:Flog} confirm this behaviour. The lessons
of this section are that
(i) arbitrary resummations may lead to misleading results, and
(ii) even if a sequence of approximations converges, it may
converge towards the wrong result.

\subsection{From moment expansion to density of states}

In the previous section we realized that the cumulant expansion
for ${\tilde D}_{\W}(\omega)$ has a finite radius of convergence,
$\omega_0\approx2.3$. Here we shall first show that the
moment expansion for $D_{\W}(\omega)$,
cf.\ equation (\ref{FTWegner}), has an infinite radius of convergence.
To this end we must evaluate $\langle \epsilon^n \rangle$ as
$n\to\infty$. This quantity will receive its main contribution
from the region of large $\vert\epsilon\vert$, where we may
approximate ${\tilde D}_{\W}(\epsilon)\approx 2 \pi^{-1/2}\, \epsilon^2 \,
\exp(-\epsilon^2)$. An asymptotic evaluation of the
integral $\int_{-\infty}^{\infty}\d\epsilon\,
\epsilon^{n+2} \e^{-\epsilon^2}$ then reveals that it receives
its main contribution from $\vert\epsilon\vert\approx\sqrt{1+n/2}$ (confirming
the approximation), and that
\[
    \langle \epsilon^n \rangle \sim \frac{2}{\sqrt{\pi}}\,
    \Gamma\left(\frac{n+3}{2}\right)\quad \mbox{as $n\to\infty$}.
\]
It follows that the series ${\tilde D}_{\W}(\omega)=
\sum_{n=0}^{\infty} (-\omega^2)^n\,
\langle\epsilon^{2n}\rangle/2n!$ converges for all $\omega$. Clearly,
a term-by-term integration of this series will not lead to a sensible
density of states. However, we may extract a convergence factor
$\e^{-a\,\omega^2/2}$, and series expand the quantity
$\e^{a\,\omega^2/2}\,{\tilde D}_{\W}(\omega)$, which also has an infinite
radius of convergence. With (the somewhat natural choice of) $a=1$ this
method gives better results than the cumulant expansion,
but the convergence with $n$ is unimpressive.
(With {\em a priori} knowledge of the answer it is indeed
possible to choose the parameter $a$ such that an excellent reconstruction
of the density of states is obtained. However, we have not found a good
objective criterion for choosing $a$, which would work in
more general circumstances.)

\pagebreak[4]

\subsection{The known moments as constraints on the density of states}

\begin{figure}[ht]
\hbox{
\hspace{0.16\textwidth}\epsfig{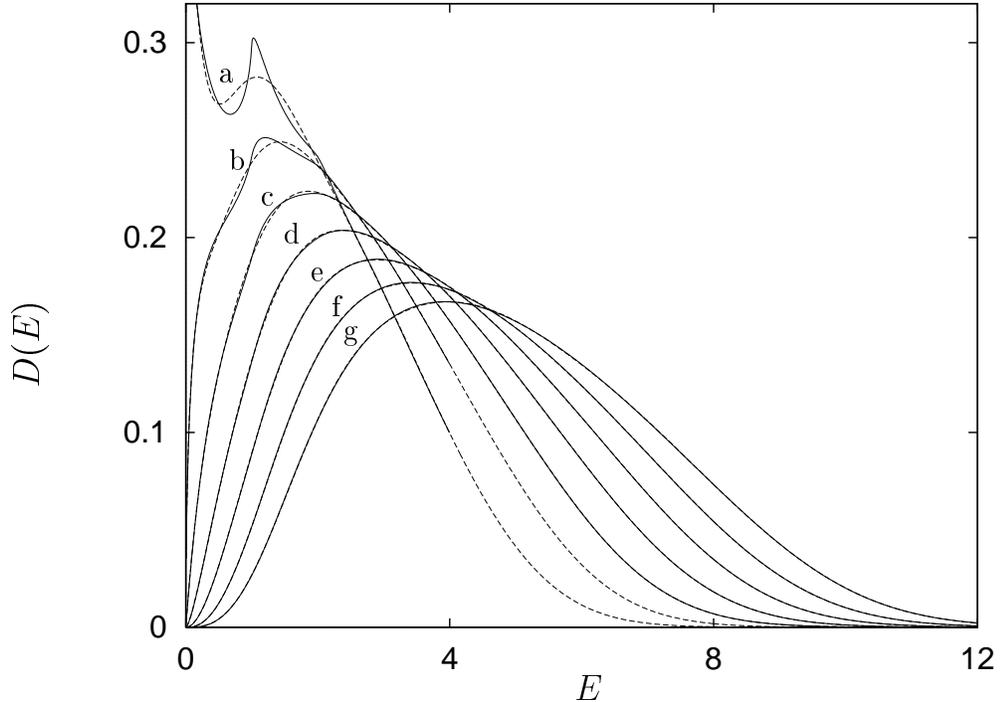}
}
\caption{A comparison of the exact density of states,
$D_{\B}(E)$ (fully drawn lines) and the density $D(E)$ reconstructed
from the 12 first moments (dashed lines) for
a) $f=1.99$, b) $f=2.5$, c) $f=3.0$, d) $f=3.5$, e) $f=4.0$,
g) $f=4.5$, and h) $f=5.0$.
}
\label{fig:Fall}
\end{figure}

\begin{figure}[ht]
\hbox{
\hspace{0.16\textwidth}\epsfig{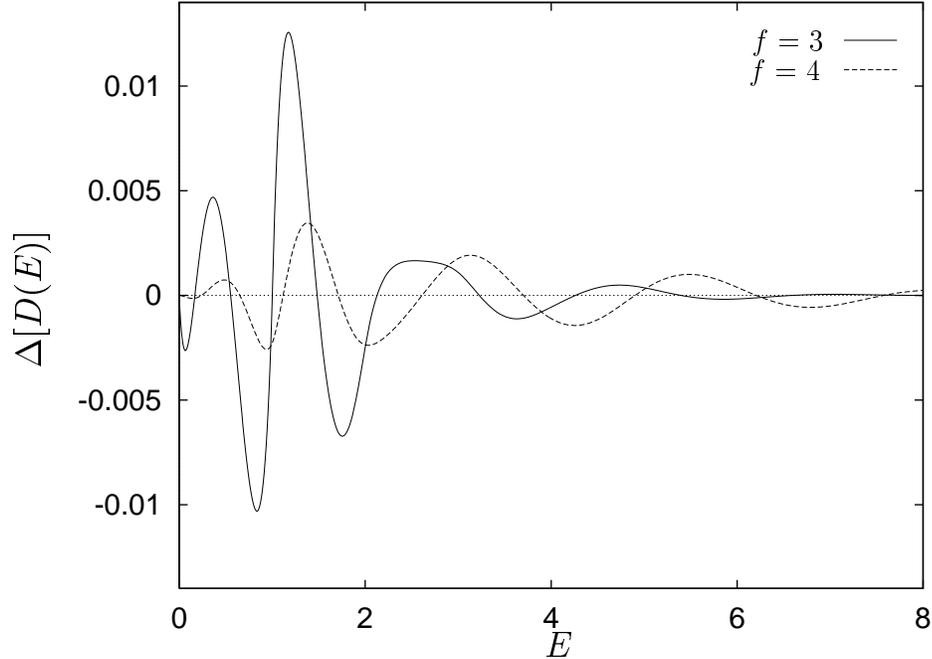}
}
\caption{The accuracy of the reconstructed density. We plot the quantity
$(D(E)-D_{\B}(E))/D_{\B}^{(max)}$ for $f=3$ (fully drawn line) and
$f=4$ (dashed line), where $D_{\B}^{(max)}$ is the maximal value of the
function $D_{\B}(E)$.}
\label{fig:Fdiff}
\end{figure}

Given only a finite set of moments,  $\langle E^k \rangle$ for $k=0,\ldots,n$,
any non-negative function $D(E)$ which reproduces them is in principle
a possible solution for the corresponding density of states. In practice,
$D(E)$ may be expected to have some simple asymptotic behaviour
for large $E$. Since $\langle E^k \rangle$ for large $k$ is mainly
determined by $D(E)$ for large $E$, we may use the asymptotics of
$\langle E^k \rangle$ to estimate this tail of $D(E)$. Further, for
the models at hand, we know that $D(E)=0$ for $E<0$, and we may have
some independent information about how $D(E)$ behaves at small $E$.
With the behaviour of $D(E)$ constrained from both sides, one
expect it to be well determined by its lowest moments---provided it
is a reasonably smooth function.

This strategy gives a satisfactory method to
reconstruct the density of states.
We find that the asymptotics of our calculated moments fits well
to the behaviour
\[
     \frac{\langle E^k \rangle}{\langle E^{k-1}\rangle} \sim
     \alpha k^{1/2} + \beta + {\cal O}(k^{-1/2})
\]
Such a behaviour is reproduced by a density of states which behaves like
\[
     D(E) \sim   E^{c}\,\exp\left(-a E^2 + b E\right)
     \quad
     \mbox{as $E\to\infty$.}
\]
We have the connection $\alpha=(2a)^{-1/2}$ and $\beta=b (2a)^{-1}$. The index
$c$ does not enter to the order we consider here. With the coefficients $a$ and $b$
determined we write the density of states in the form
\be
     D(E) = E^{f-2}\,\exp\left(-aE^2+bE\right)\times P_n(E),
\ee
where $P_n$ is a $n$'th order polynomial. We determine its
$n+1$ coefficients such that all the moments
$\langle E^0 \rangle,\ldots\langle E^n \rangle$ are reproduced.
This construction leads to a $D(E)$ which compares favourably
with the corresponding $D_{\B}(E)$,
at least when $f$ is somewhat larger than $2$, cf.\ figure~\ref{fig:Fall}.

Since the difference between the exact and the reconstructed curves
is mostly invisible in figure~\ref{fig:Fall}, we show some examples of this
difference in figure~\ref{fig:Fdiff}. The accuracy is seen to improve
with increasing $f$,
most likely because the function we try to reconstruct becomes smoother
(thus this trend may not continue to
arbitrary high $f$). In this reconstruction
we have built in the known low-$E$ behaviour,
$D(E)\sim E^{f-2}$. We have noted that the overall
reconstruction is fairly insensitive to shifting
the exponent away from $f-2$.
Choosing the correct exponent may give
somewhat better convergence with $n$.


%% file: rho.tex
\section{Density of states for general
{\protect\boldmath $\varrho$}\label{sec:rho}}

\begin{figure}[ht]
\hbox{
\hspace{0.16\textwidth}\epsfig{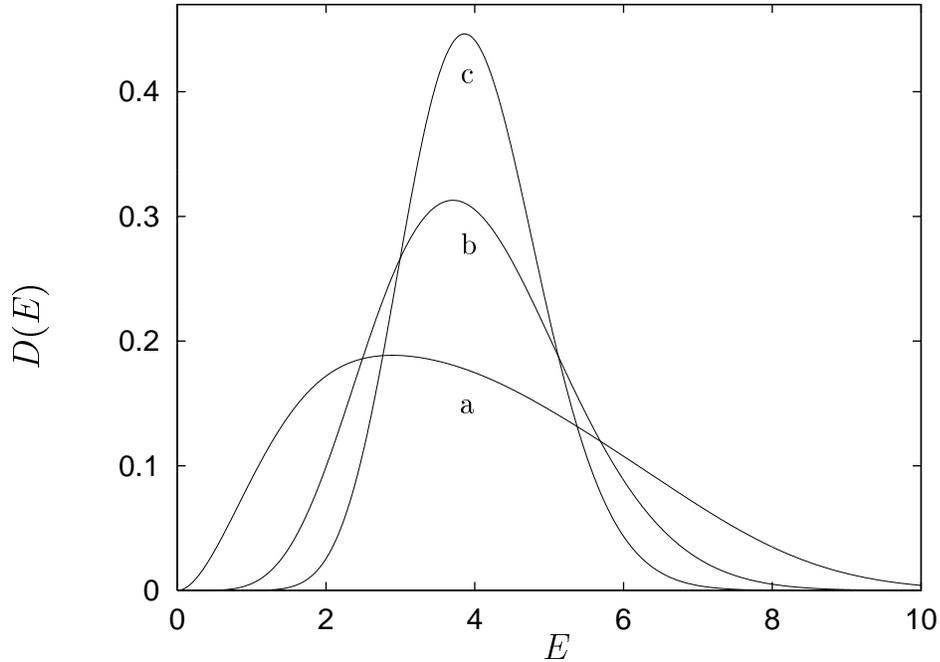}
}
\caption{Constructed density of states for $f=4$ and a) $\varrho=1.0$,
b) $\varrho=2.5$, and c) $\varrho=5.0$.}
\label{fig:Ffrange}
\end{figure}

The method used in the previous subsection can be used equally well
to reconstruct the density of states for $\varrho>1$. The main difference
is that we have no {\em a priori} knowledge of the low-$E$ behaviour of
$D(E)$. However, in the previous section we gave a heuristic
probability argument for the $f$-dependence of the exponent.
This argument may be repeated for a potential of finite range.
It suggests that one should make the replacement
\be
       f \to {\bar f} = \case{1}{2}(1+\varrho) f
\ee
in the exponent when $\varrho>1$. Thus,
we have imposed the requirement that
$D(E)\sim E^{{\bar f}-2}$ as $E\to0^+$. The resulting density
of states is shown in figure~\ref{fig:Ffrange} for
a set of $\varrho$-values. We now longer have exact results to
compare against, but comparison with numerical simulations show
excellent agreement. We believe the difference from the exact curves would
not be visible in the plots.

As $\varrho$ increases with $f$ fixed, the distribution approaches a Gaussian
of width $\sqrt{f/\varrho}$, centred around $E=f$:
\be
  D(E) \approx (2\pi f/\varrho)^{-1/2}
  \exp\left(-\frac{1}{2} \frac{(E-f)^2}{f/\varrho}\right).
\ee
In terms of the variable $\epsilon\equiv(E-f)\sqrt{\varrho/f}$, the first
high-$\varrho$ correction to this distribution is determined by the third
moment
\be
   \langle \epsilon^3 \rangle \approx \case43 (\varrho f)^{-1/2}. 
\ee
In general, the $k$'th order correction (as e.g.\ defined by the $k$'th cumulant
in $\epsilon$) vanishes
like $\varrho^{(2-k)/2}$ as $\varrho\to\infty$.


%% file: higher.tex
\section{Higher Landau levels \label{sec:higher}}

Calculation of averaged Green functions in the higher Landau levels
proceed as in the lowest one. The changes are that we measure energy
relative to the band centre, $E=(\nu+1/2)$, and use the appropriate
expression for the projection operator, i.e. $\cP \rightarrow \cP_{\nu}$,
where $\nu$ is the Landau level index. The (integration kernel) of
the projection operator for arbitrary Landau level is given in \ref{app:P}.
The perturbation expansion for the averaged Green function continues
to be equivalent to computing the moments of the energy density of states.
Since we shall only perform a low order calculation ``by hand'' we may
as well calculate the moments directly. (The main advantage of working
with the Green function instead of directly with the moments is that
we don't have to worry about combinatorial factors. This reduces the
time it takes to code, debug and run computer algebra routines.)

\subsection{Evaluation of the three lowest moments}

\begin{figure}[ht]
\hbox{
\hspace{0.16\textwidth}\epsfig{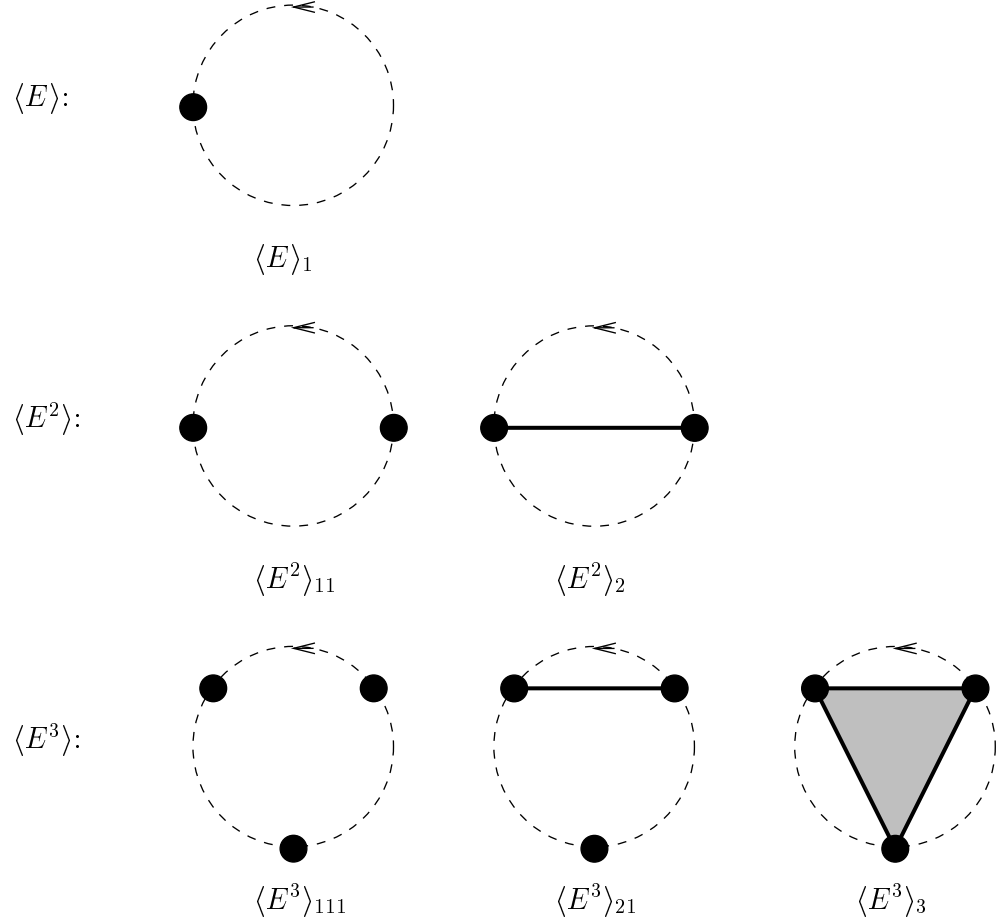}
}
\caption{Graphs constituting the first three energy moments
$\langle E\rangle,\,
\langle E^2\rangle$ and $\langle E^3\rangle$.
Note that the statistical weight is a factor $k!$
higher than the usual combinatorial factor, where $k$
is the order of the moment considered.
}
\label{fig:Fgraph1}
\end{figure}

The diagrams for the first three energy moments
are shown in figure~\ref{fig:Fgraph1}.

\begin{figure}[ht]
\hbox{
\hspace{0.16\textwidth}\epsfig{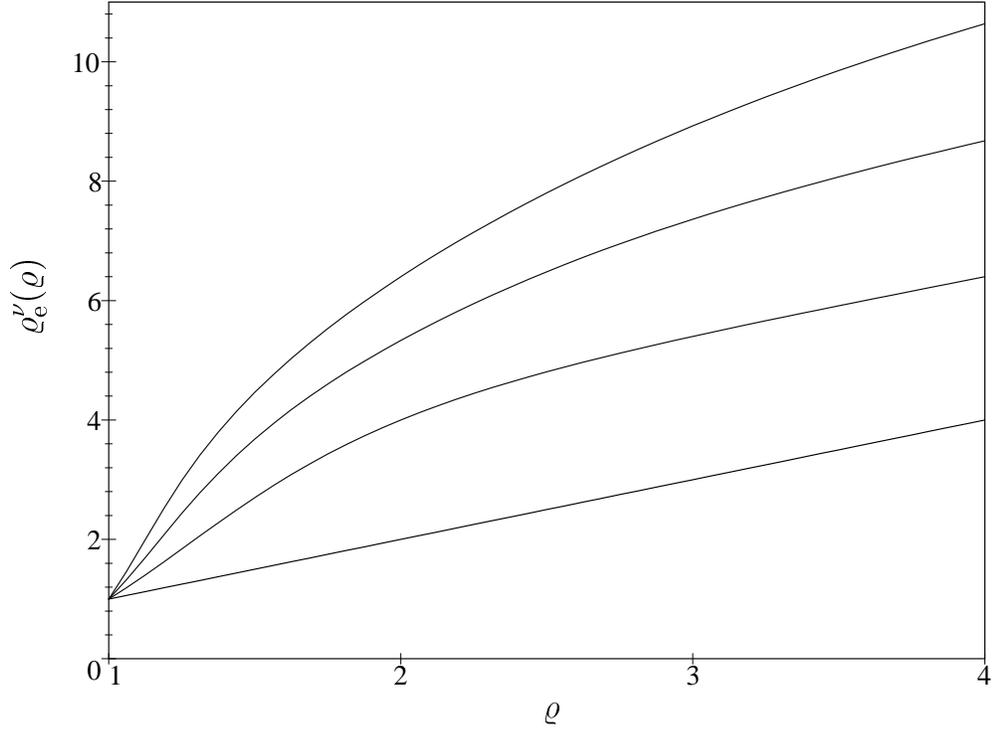}
}
\caption{The rescaled, effective measure of impurity width, 
$\varrho_{\e}^{\nu}$ plotted as a function of $\varrho$,
for the four lowest Landau levels, for $\nu=0,1,2,3$.
The curve is shifted upwards for increasing $\nu$.}
\label{fig:Frhoeff}
\end{figure}

We have
\be
   \langle E \rangle_1=f,\quad
   \langle E^2 \rangle_{11}=f^2,\quad
   \langle E^3 \rangle_{111}=f^3,\quad
   \langle E^3 \rangle_{12}=3f\,\langle E^2 \rangle_{2},  
\ee
which reduces the problem to evaluating $\langle E^2 \rangle_{2}$
and $\langle E^3 \rangle_{12}$. These quantities depend on the Landau
level index, which we shall indicate in our notation below.
The evaluation of $\langle E^2 \rangle_{2}$ reduces to computing
\[
  \langle E^2\rangle_{2}^{(\nu)}=\frac{f}{\varrho}
  \itot \frac{\d {\vec u}}{2\pi}
  L_{\nu}^{2}(\,\case12(1-\case1\varrho) {\vec u}^{\,2}\,) \,
  \exp\left(-\case12 {\vec u}^{\,2}\right),
\]
where $L_{\nu}$ is the $\nu$'th Laguerre polynomial. Since this is a Gaussian times
a polynomial, it is simple to evaluate for the first few $\nu$. We find
\[
\begin{array}{cc} 
\hline 
\nu & \langle E^2\rangle_{2}^{(\nu)} \rule{0mm}{5mm} \\
\hline \hline
0 & \frac{1}{\varrho} f \rule{0mm}{5mm} \\
1 & \frac{(\varrho-1)^2+1}{\varrho^{3}} f \rule{0mm}{5mm} \\
2 & \frac{(\varrho-1)^4+4 (\varrho-1)^2+1}{\varrho^{5}} f \rule{0mm}{5mm} \\
3 & \frac{(\varrho-1)^6+9 (\varrho-1)^4+9 (\varrho-1)^2+1}{\varrho^{7}} f \rule{0mm}{5mm} \\
\ms
\hline
\end{array}
\]
This fits with a general expression
\bea
  \langle
     E^2\rangle_{2}^{(\nu)}=
     f\,{\varrho^{-(2\nu+1)}}
     \sum_{k=0}^{\nu} \left[ {\nu \choose k} (\varrho-1)^k \right]^2
     \equiv\frac{f}{\varrho_{\e}^{(\nu)}},
  \label{eq:E2nu}
\eea
where $\varrho_{\e}^{(\nu)}$ is a rescaled, effective form of $\varrho$, allowing us to write
the second order moment for arbitrary $\nu$ in the same form as the second 
order moment for the lowest Landau level. We show in figure~\ref{fig:Frhoeff} how this
quantity varies with $\rho$ and $\nu$.

A similar evaluation of $\langle E^3 \rangle_{3}^{(\nu)}$
for the first few $\nu$ gives
\[\fl
\begin{array}{cc} 
\hline 
\nu & \langle E^3 \rangle_{3}^{(\nu)} \rule{0mm}{5mm} \\
\hline \hline
0 &  \frac{4}{1+3 \varrho^2} f \rule{0mm}{5mm} \\
1 & \frac{4}{(1+3 \varrho^2)^{4}} \left( 27 {\varrho}^{6}-108 {\varrho}^{5}+207{\varrho}^{4}-168{\varrho}^{3}
                                     +33{\varrho}^{2}+84{\varrho}-11\right) f \rule{0mm}{5mm} \\
2 & \frac{4}{(1+3 \varrho^2)^{7}}
         \left( 729{\varrho}^{12}-5832{\varrho}^{11}+24786{\varrho}^{10}-61560{\varrho}^{9}
                                    +94527{\varrho}^{8}-72144{\varrho}^{7} \right. \rule{0mm}{5mm} \\
  &   \left. -23076{\varrho}^{6}+114192{\varrho}^{5}
                                   -104697{\varrho}^{4}+29976{\varrho}^{3}+11826 {\varrho}^{2}-4632 {\varrho}+1
                                  \right)f \rule{0mm}{5mm} \\
\ms
\hline
\end{array}
\]
Here the generalization to arbitrary $\nu$ is not apparent, except that
specializing to $\varrho=1$ gives $\langle E^3 \rangle_3^{(\nu)}=f$.
The expressions above are related to the cumulants by
\be
  \chi_{n}^{(\nu)} = \frac{(-1)^{n}}{n!}\,
   \langle E^n \rangle^{(\nu)}_{n}\qquad\mbox{for $n=1,2,3$.}
\ee
(This relation does not generalize to higher $n$.)

\pagebreak[4]

\subsection{From moments to density of states}

\begin{figure}[ht]
\hbox{
\hspace{0.16\textwidth}\epsfig{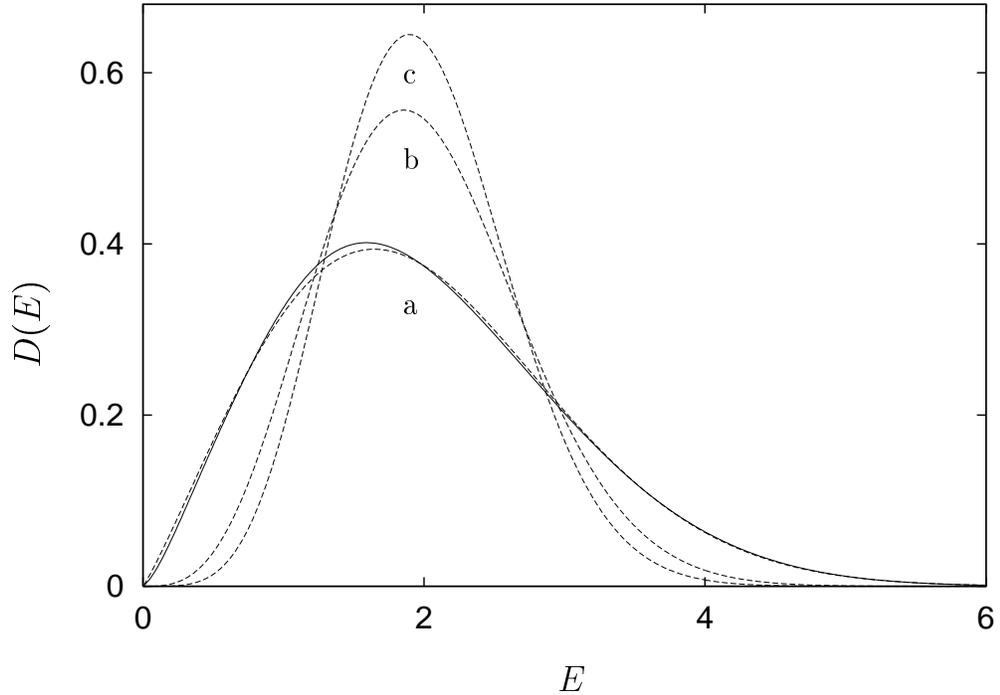}
}
\caption{Density of states constructed from
$\langle E\rangle$, $\langle E^2 \rangle$, and $\langle E^3\rangle$
(dashed lines) for a) $\nu=0$, b) $\nu=1$, and c) $\nu=2$, with
$f=\varrho=2$. For comparison we show the density constructed from
the 12 first moments in the lowest Landau level (full line).
}
\label{fig:Fhigher}
\end{figure}

In this case we have calculated too few moments to use the method
from sections~5--6. Instead we parametrize the density of states as
\be
    D(E) = C\,E^c\,\exp\left(-aE^2\right)\left(1+b E \right),
\ee
where $C$ is a normalization constant, and the parameters $a$, $b$, $c$
are chosen such that $\langle E \rangle$, $\langle E^2 \rangle$, and
$\langle E^3 \rangle$ are reproduced. To get some measure of the
accuracy obtained, we also use the same procedure for the lowest Landau level.
For $f=\varrho=2$ we find the following fitting parameters
\[
\begin{array}{cccc} 
\hline 
\nu & a & b & c \rule{0mm}{5mm} \\
\hline \hline
0 & 0.22174 & 0.06732 & 1.10221 \rule{0mm}{5mm} \\
1 & 0.47681 & 0.22997 & 2.98668 \rule{0mm}{5mm} \\
2 & 0.64008 & 2.50640 & 3.77993 \rule{0mm}{5mm} \\
\ms
\hline
\end{array}
\]
The resulting densities are plotted in figure~\ref{fig:Fhigher}.
For comparison we also show the density found for the lowest
Landau level when all 12 moments are used.
It seems that $D(E)$ can be reproduced fairly well from only
three moments.


%% file: discussion.tex

\section{Concluding Remarks.\label{sec: discussion}}

We have in this paper generated and analysed
a high ($12$'th) order perturbation expansion of
the averaged Green function, in a class of
models inspired by the integer Quantum Hall
Effect. The generation process can be fully
automatized using computer algebra
programs. The order which can be obtained
by our programs
is mostly limited by available CPU time,
as the number of diagrams exhibit factorial
growth with the perturbative order. With present
technology about 3 more orders would be feasible
for the quantity considered.

For the models considered the $n$'th order
perturbation expansion provides exact information
of the $n$ first moments of the density of states
versus energy, $D(E)$. When $D(E)$ is
a reasonably smooth function it can be reconstructed
to high accuracy (better than 1 percent)
from the computed moments. The results are also
useful to check and complement other approaches,
like numerical simulations on finite size systems.

There are several directions to extend our method. The one
of most immediate interest for the integer Quantum
Hall Effect is to extend the analysis to the averaged
2-particle Green function. This quantity encode
information about transport properties.
We have done some initial investigations in this direction.
It is fairly straightforward to implement a
computer algebra procedure which automatically generate
all diagrams, and evaluates them. The number of diagrams
will be about one order of magnitude larger (for e.g.\ the
$12$'th perturbative order). The information generated can be viewed
as various moments of a 3-variable spectral function.
It is not clear how easy it will be to reconstruct this
spectral function, or the interesting physical information
it contains, from the information generated by the perturbation
expansion.


%% file: acknowl.tex
\ack
This work was done in part at the Center for Advanced Study,
Oslo, Norway. We thank the center for kind hospitality
and financial support.
The work of one of us (A.K.) has been partially funded by the
Norwegian Research Council under contract no. 100559/410.

%% file: appA.tex
\section{Projection operator onto the
\mbox{\protect\boldmath $\nu$'th} Landau level \label{app:P}}

With appropriate units and gauge choice, the imaginary time propagator
for a $2D$ electron (with charge $-e$, where $e>0$)
in a magnetic field solves the initial value
problem\footnote{
Here $\vec r$ and $\vec\rho$ are $2D$ vectors with components
$(x,y)$ and $(\xi,\eta)$ respectively.
Units for length and time are chosen
such that $\ell_B \equiv (\hbar/e B)^{1/2}$ and
$\omega_B \equiv e B/m$ equals unity.
Orientation is chosen such that $\vec B$ points along the positive
$z$-axis, and the problem is formulated in the (radial) symmetric gauge,
${\vec r}\cdot{\vec A}({\vec r}) = 0$.
}
\begin{eqnarray}
  \frac{\partial}{\partial\tau}\,G({\vec r},{\vec \rho};\tau) &=&
  \frac{1}{2}\left[\left(\frac{\partial}{\partial x} -\frac{\i}{2}y\right)^2+
  \left(\frac{\partial}{\partial y} +\frac{\i}{2}x\right)^2 \right]
  \,G({\vec r},{\vec \rho};\tau)\\
  G({\vec r},{\vec\rho};0) &=& \delta({\vec r}-{\vec\rho})
\end{eqnarray}
By making the ansatz that $G$ is a Gaussian in $(x,y,\xi,\eta)$,
with prefactor and coefficients which depend on $\tau$,
one obtains the solution
\begin{equation}
  G({\vec r},{\vec\rho};\tau) = \frac{\sqrt{\varepsilon}}{2\pi}\,
  \mbox{e}^{\i\left(x \eta-\xi y\right)/2}
  \frac{1}{1-\varepsilon}\,
  \mbox{e}^{-\varepsilon R/(1-\varepsilon)}\;\mbox{e}^{-R/2},
  \label{propagator}
\end{equation}
where $\varepsilon=\mbox{e}^{-\tau}$ and
$R=\frac{1}{2}({\vec r}-{\vec \rho})^2$.
Since the energy of the $\nu$'th Landau level is $\nu+\frac{1}{2}$,
$\nu=0,1,\ldots$, an alternative expression for $G$ is
\begin{equation}
  G({\vec r},{\vec\rho};\tau) = \mbox{e}^{-\tau/2}\,
  \sum_{\nu=0}^{\infty}\;{\cal P}_{\nu}({\vec r},{\vec\rho})\,\mbox{e}^{-\nu\tau},
\end{equation}
where ${\cal P}_{\nu}$ is the projection onto the
$\nu$'th Landau level.
Thus, by series expanding equation~\eref{propagator} in powers of
$\varepsilon$ we easily find the (integral kernels of the)
projection operators onto the various Landau levels. The first
few examples are
\begin{eqnarray}
  {\cal P}_0({\vec r},{\vec \rho}) &=&
  \frac{1}{2\pi}\,\mbox{e}^{\i(x \eta-\xi y)/2}\;
  \mbox{e}^{-R/2},
  \\
  {\cal P}_1({\vec r},{\vec \rho}) &=&
  \frac{1}{2\pi}\,\mbox{e}^{\i(x \eta-\xi y)/2}\;
  (1-R)\,\mbox{e}^{-R/2},
  \\
  {\cal P}_2({\vec r},{\vec \rho}) &=&
  \frac{1}{2\pi}\,\mbox{e}^{\i(x \eta-\xi y)/2}\;
  (1-2 R + \case12 R^2)\,\mbox{e}^{-R/2}.
  \label{eq:Ps}
\end{eqnarray}
The general expression is
\begin{equation}
  {\cal P}_{\nu}({\vec r},{\vec \rho}) =
  \frac{1}{2\pi}\,\mbox{e}^{\i(x \eta-\xi y)/2}\;
  L_{\nu}(R)\,\mbox{e}^{-R/2},  
\end{equation}
with $L_{\nu}$ the $\nu$'th Laguerre polynomial.

\subsection{Eigenvalues in a rotation symmetric potential}

The projection operators can be expanded in a basis of rotation
eigenfunctions. With $x=r\cos\varphi$, $y=r\sin\varphi$,
$\xi=r'\cos\varphi'$, $\eta=r'\sin\varphi'$: 
\be
   P_{\nu}({\vec r},{\vec \rho}) = 
   \sum_{\ell=-\nu}^{\infty} \e^{-\i\ell(\varphi-\varphi')}\,
   F^{(\nu)}_\ell(r)\,F^{(\nu)}_\ell(r').
   \label{RotationBasis}
\ee
If we introduce a rotation symmetric potential,
$W({\vec r})=W(r)$, $\{\, \e^{-\i\ell\varphi}\,F_\ell(r)\,\}$
continues to be a basis of energy eigenfunctions.
The corresponding eigenvalues are
\be
   E^{(\nu)}_{\ell}= 2\pi\,\int_0^\infty r\d r\,
   F^{(\nu)}_{\ell}(r)^2\,W(r).
   \label{EigenEnergies}
\ee
Combining (\ref{RotationBasis}-\ref{EigenEnergies}) gives
a generating function for the eigenvalues,
\be
    Z^{(\nu)}(\e^{-i\varphi})=2\pi\,\int_0^{\infty} r\d r\,
    P_{\nu}({\vec r},{\vec\rho})\,W(r) =
    \sum_{\ell=-\nu}^{\infty}\,E^{(\nu)}_{\ell}\,\e^{-\i\ell\varphi}.
\ee
Here ${\vec\rho}$ is chosen such that $r'=r$ and $\varphi'=0$.
With $W(r)$ as defined in equation
(\ref{OneImpurityPotential}) the integration is simple, and we
get a generating function
\be
    Z^{(\nu)}(\e^{-i\varphi})= \frac{2}{\varrho+1}\,
    (a \e^{\i\varphi})^\nu\,
    (1-b\e^{-\i\varphi})^\nu\,(1-a\e^{-\i\varphi})^{-\nu-1},
\ee
where $a=(\varrho-1)/(\varrho+1)$ and $b=(\varrho-3)/(\varrho-1)$.
It is easy to verify that
\[
 Z^{(\nu)}(1)= \sum_{\ell=-\nu}^\infty\,E^{(\nu)}_\ell=1.
\]
Series expansion gives the eigenvalues
\be
   E^{(\nu)}_{\ell} = (-1)^{\ell+\nu}\,\frac{2}{\varrho+1}\,a^{\ell+2\nu}\,
   \sum_{k=0}^{\nu} {\nu \choose k}\,{-\nu-1 \choose \nu+\ell-k}\,
   \left(\frac{b}{a}\right)^k.
\ee
We find that $E^{(\nu)}_{\ell}\to\delta_{\ell,0}$ as $\varrho\to1^+$.
More explicit expressions for the first three Landau levels are
\bea
   E^{(0)}_{\ell} = \frac{2}{\varrho+1}\,
   \left(\frac{\varrho-1}{\varrho+1}\right)^{\ell},\\
   E^{(1)}_{\ell} = \frac{2}{(\varrho+1)^3}\,
   \left(\frac{\varrho-1}{\varrho+1}\right)^{\ell}
   \left[4(\ell+1)+(\varrho-1)^2\right],\\
   E^{(2)}_{\ell} = \frac{2}{(\varrho+1)^5}\,
   \left(\frac{\varrho-1}{\varrho+1}\right)^{\ell}\times\nonumber\\
   \left[8(\ell+2)(\ell+1)+8(\ell+2)(\varrho-1)^2+(\varrho-1)^4\right].
\eea

%% file: appC.tex

\section{Tables of cumulants.\label{app:CumTabs}}

We here present the 12 first cumulants for some selected values
of $\varrho$. The exact expressions are much too long to present
in full (except when $\varrho=1$), thus we give only their numerical
approximations. It is apparent from these tables that the higher
cumulants vanishes quite rapidly when $\varrho$ is large (and
$f$ is of order one).

\begin{table}
\caption{The 12 first cumulants for $\varrho=1$.}
\begin{indented}
\item[]\begin{tabular}{@{}rl}\br
$k$& $(-1)^k \, k! \, \chi_k$ \\ 
 \mr
1 &  $f$ \\ \ms
2  & $f $ \\ \ms
3  & $f $ \\ \ms
4  & $f -0.500000  f^2 $ \\ \ms
5  & $f -2.500000  f^2 $ \\ \ms
6  & $f -8.166667  f^2 +1.750000  f^3 $ \\ \ms
7  & $f -22.16667  f^2 +19.25000  f^3 $ \\ \ms
8  & $f -54.41667  f^2 +126.1667  f^3 -13.625000  f^4 $ \\ \ms
9  & $f -125.7500  f^2 +644.6111  f^3 -260.62500  f^4 $ \\ \ms
10  & $f -279.5500  f^2 +2843.403  f^3 -2844.1667  f^4 +185.687500  f^5 $ \\ \ms
11  & $f -605.5500  f^2 +11399.82  f^3 -23374.694  f^4 +5449.81250  f^5 $ \\ \ms
12  & $f -1288.383  f^2 +42793.33  f^3 -161291.55  f^4 +88986.8125  f^5 -3891.03125  f^6 $ \\ \ms
\br
\end{tabular}
\end{indented}
\label{tab:chi1}
\end{table}

\begin{table}
\caption{The 12 first cumulants for $\varrho=2$.}
\begin{indented}
\item[]\begin{tabular}{@{}rl}\br
$k$& $(-1)^k \, k! \, \chi_k$ \\ 
 \mr
1 &  $f$ \\ \ms
2 & $ 0.500000 f$ \\ \ms
3 & $ 0.307692 f$ \\ \ms
4 & $ 0.200000 f-0.050000  f^2 $ \\ \ms
5 & $ 0.132231 f-0.180995  f^2 $ \\ \ms
6 & $ 0.087912 f-0.424981  f^2 +0.046429  f^3 $ \\ \ms
7 & $ 0.058554 f-0.820875  f^2 +0.387161  f^3 $ \\ \ms
8 & $ 0.039024 f-1.420263  f^2 +1.889714  f^3 -0.095141  f^4 $ \\ \ms
9 & $ 0.026014 f-2.293572  f^2 +7.080389  f^3 -1.435478  f^4 $ \\ \ms
10 & $ 0.017342 f-3.537360  f^2 +22.60912  f^3 -12.021342  f^4 +0.349690  f^5 $ \\ \ms
11 & $ 0.011561 f-5.283592  f^2 +64.91179  f^3 -74.270396  f^4 +8.293927  f^5 $ \\ \ms
12 & $ 0.007707 f-7.711882  f^2 +172.9145  f^3 -379.11871  f^4 +106.0533  f^5 -2.01157  f^6 $ \\ \ms
\br
\end{tabular}
\end{indented}
\label{tab:chi2}
\end{table}

\begin{table}
\caption{The 12 first cumulants for $\varrho=3$.}
\begin{indented}
\item[]\begin{tabular}{@{}rl}\br
$k$& $(-1)^k \, k! \, \chi_k$ \\ 
 \mr
1 &  $f$ \\ \ms
2 & $ 0.333333 f$ \\ \ms
3 & $ 0.142857 f$ \\ \ms
4 & $ 0.066667 f-0.011111  f^2 $ \\ \ms
5 & $ 0.032258 f-0.029762  f^2 $ \\ \ms
6 & $ 0.015873 f-0.052596  f^2 +0.003872  f^3 $ \\ \ms
7 & $ 0.007874 f-0.077199  f^2 +0.024758  f^3 $ \\ \ms
8 & $ 0.003922 f-0.102003  f^2 +0.093423  f^3 -0.002627  f^4 $ \\ \ms
9 & $ 0.001957 f-0.126065  f^2 +0.271598  f^3 -0.032822  f^4 $ \\ \ms
10 & $ 0.000978 f-0.148870  f^2 +0.673638  f^3 -0.223264  f^4 +0.002953  f^5 $ \\ \ms
11 & $ 0.000489 f-0.170176  f^2 +1.501653  f^3 -1.107501  f^4 +0.064580  f^5 $ \\ \ms
12 & $ 0.000244 f-0.189903  f^2 +3.102167  f^3 -4.502556  f^4 +0.714094  f^5 -0.004752  f^6 $ \\ \ms
\br
\end{tabular}
\end{indented}
\label{tab:chi3}
\end{table}

\begin{table}
\caption{The 12 first cumulants for $\varrho=4$.}
\begin{indented}
\item[]\begin{tabular}{@{}rl}\br
$k$& $(-1)^k \, k! \, \chi_k$ \\ 
 \mr
1 &  $f$ \\ \ms
2 & $ 0.250000 f$ \\ \ms
3 & $ 0.081633 f$ \\ \ms
4 & $ 0.029412 f-0.003676  f^2 $ \\ \ms
5 & $ 0.011103 f-0.007701  f^2 $ \\ \ms
6 & $ 0.004296 f-0.010755  f^2 +0.000597  f^3 $ \\ \ms
7 & $ 0.001686 f-0.012572  f^2 +0.003044  f^3 $ \\ \ms
8 & $ 0.000667 f-0.013302  f^2 +0.009236  f^3 -0.000136  f^4 $ \\ \ms
9 & $ 0.000265 f-0.013214  f^2 +0.021700  f^3 -0.001567  f^4 $ \\ \ms
10 & $ 0.000105 f-0.012575  f^2 +0.043644  f^3 -0.009292  f^4 +0.000004  f^5 $ \\ \ms
11 & $ 0.000042 f-0.011601  f^2 +0.079054  f^3 -0.039239  f^4 +0.000752  f^5 $ \\ \ms
12 & $ 0.000017 f-0.010458  f^2 +0.132857  f^3 -0.134113  f^4 +0.010015  f^5 +0.000100  f^6 $ \\ \ms
\br
\end{tabular}
\end{indented}
\label{tab:chi4}
\end{table}
